\documentclass[twoside,11pt]{article}
\usepackage{graphicx}
\usepackage{bm}
\usepackage{float}
\usepackage[round]{natbib}
\usepackage{booktabs}
\usepackage{lipsum}
\usepackage{cite}
\usepackage{appendix}
\usepackage[margin=1in]{geometry}
\usepackage[usenames]{color}
\usepackage{amsfonts, amssymb, amsmath, amsthm, multicol}
\usepackage[colorlinks=true, pdfstartview=FitV, linkcolor=blue,
citecolor=blue, urlcolor=blue]{hyperref}
\usepackage[usenames]{color}
\usepackage[bottom]{footmisc}
\usepackage{graphicx}%
\usepackage{multirow}%
\usepackage{amsmath,amssymb,amsfonts}%
\usepackage{amsthm}%
\usepackage{mathrsfs}%
\usepackage{xcolor}%
\usepackage{textcomp}%
\usepackage{manyfoot}%
\usepackage{booktabs}%
\usepackage{algorithm}%
\usepackage{algorithmicx}%
\usepackage{algpseudocode}%
\usepackage{listings}%

\usepackage[figuresright]{rotating}%
\usepackage{threeparttable}
\usepackage{lscape}
\usepackage{soul}
\usepackage[section]{placeins}

\usepackage{tikz}
\usetikzlibrary{shapes, arrows}
\usepackage{authblk}
\definecolor{Red}{rgb}{1,0.05,0}
\definecolor{Grn}{rgb}{0.1,0.7,0.1}
\definecolor{Blu}{rgb}{0.1,0.1,0.6}
\definecolor{Org}{rgb}{1,0.45,0}
\definecolor{Vio}{rgb}{0.6578,0,0.9478}
\definecolor{Mag}{rgb}{1,0.2,0.3}
\title{Error propagation dynamics of velocimetry-based pressure field calculations (4): on the impact of flow profile}

 \author[1]{\textcolor{Grn}{Lanyu Li}}
 \author[1]{\textcolor{Red}{Zhao Pan}\thanks{To whom correspondence may be addressed: zhao.pan@uwaterloo.ca}}
  \affil[1]{Dept. of Mechanical and Mechatronics Engineering, University of Waterloo, ON, Canada}

\date{ }

\title{Three-Dimensional Time Resolved Lagrangian Flow Field Reconstruction Based on Constrained Least Squares and Stable Radial Basis Function}

\begin{document}

\maketitle

\begin{abstract}
The three-dimensional Time-Resolved Lagrangian Particle Tracking (3D TR-LPT) technique has recently advanced flow diagnostics by providing high spatiotemporal resolution measurements under the  Lagrangian framework.
To fully exploit its potential, accurate and robust data processing algorithms are needed.
These algorithms are responsible for reconstructing particle trajectories, velocities, and differential quantities (e.g., pressure gradients, strain- and rotation-rate tensors, and coherent structures) from raw LPT data.
In this paper, we propose a three-dimensional (3D) divergence-free Lagrangian reconstruction method, where three foundation algorithms---Constrained Least Squares (CLS), stable Radial Basis Function (RBF-QR), and Partition-of-Unity Method (PUM)---are integrated into one comprehensive reconstruction strategy.
Our method, named CLS-RBF PUM, is able to 
(i) directly reconstruct flow fields at scattered data points, avoiding Lagrangian-to-Eulerian data conversions; 
(ii) assimilate the flow diagnostics in Lagrangian and Eulerian descriptions to achieve high-accuracy flow reconstruction;
(iii) process large-scale LPT data sets with more than hundreds of thousand particles in two dimensions (2D) or 3D; 
(iv) enable spatiotemporal super-resolution while imposing physical constraints (e.g., divergence-free for incompressible flows) at arbitrary time and location.
Validation based on synthetic and experimental LPT data confirmed that our method can consistently achieve the above advantages with accuracy and robustness.  
\end{abstract}

\section{Introduction}
\label{introduction}
The high seeding density three-dimensional Time-Resolved Lagrangian Particle Tracking (3D TR-LPT) is a powerful tool for high spatiotemporal resolution flow diagnostics ~\citep{schanz2016shake,tan2020introducing}.
This technique offers three major advantages in fluid experiments.
First, the 3D TR-LPT works in the Lagrangian perspective, which is intuitive and allows for a heuristic interpretation of fluid flows.
For example, one of the most famous early Lagrangian flow `diagnostics' is perhaps tracking the movement of tea leaves in a stirred teacup, as elucidated by Einstein's study on the tea leaf paradox 
 \citep{einstein1926ursache,bowker1988albert}. 
Second, the LPT technique, including 3D TR-LPT, 
facilitates trajectory-based measurements, providing insights into the history of the flows.
By following the pathlines of tracer particles, one can recover the past and predict the future of the flows.
For example, the LPT has been applied in particle residence time studies \citep{jeronimo2019direct,zhang2020direct}, indoor airflow measurements to improve air quality \citep{biwole2009complete,fu2015particle} and fluid mixing studies \citep{alberini2017comparison,romano2021lagrangian}. 
Third, the 3D TR-LPT technique excels at capturing transient and time-correlated flow structures. 
This feature is crucial for studying complex flows characterized by intricate evolution of the flow structures (e.g., Lagrangian coherent structures \citep{peng2009transport,wilson2009lagrangian,rosi2015lagrangian,onu2015lcs} that are the `skeleton' of fluid flows \citep{peacock2013lagrangian}).

However, processing the raw LPT data is not necessarily trivial for a few reasons.
First, key flow quantities, including particle trajectories, velocities, and differential quantities (e.g., pressure gradients, strain- and rotation-rate tensors) are not directly available in raw data.
Technically, the raw LPT data only consist of particle spatial coordinates as a time series and anything beyond those requires additional reconstruction. 
Second, the particle spatial coordinates in the raw data are often subject to measurement errors \citep{ouellette2006quantitative,cierpka2013higher,gesemann2016noisy,machicoane2017multi}.
Deviations between the true and measured particle locations are inevitable and can challenge the data processing and undermine reconstruction quality. 
Furthermore, (numerical) differentiation amplifies the noise present in the data when computing derivatives, leading to even noisier reconstructions without careful regularization.
This effect is evident when reconstructing velocities and differential quantities. 
Third, the scattered and nonuniform distribution of LPT data makes it difficult to employ simple numerical schemes, such as finite difference methods, for calculating spatial derivatives.  
To address these challenges and process the raw LPT data, several methods have been proposed and we provide a brief summary below.

In the context of trajectory reconstruction, polynomials and basis splines (B-splines) are commonly used, 
with an order being second \citep{cierpka2013higher}, third \citep{luthi2002some,cierpka2013higher,gesemann2016noisy,schanz2016shake}, or fourth \citep{ferrari2008particle}.
Typically, least squares regression is applied to mitigate the impact of noise in the particle coordinates \citep{luthi2002some,cierpka2013higher,gesemann2016noisy,schanz2016shake}. 
However, polynomials and B-splines may encounter the following difficulties.
First, low-order functions cannot capture high-curvature trajectories, while high-order ones may be prone to numerical oscillations known as Runge's phenomenon \citep{gautschi2011numerical}. 
Second, developing self-adaptive schemes that can accommodate trajectories with varied curvatures is not a trivial task.
In the existing methods, the same trajectory function with a fixed order is commonly used throughout the entire reconstruction, offering limited flexibility to approximate varied curvature trajectories.
Third, achieving a high degree of smoothness with high-order polynomials becomes challenging when the number of frames is limited. 
For example, with four frames, the maximum attainable order of a polynomial is third. 
This implies that the particle acceleration varies linearly, and its jerk is a constant, which may not always be true.

Velocity reconstructions are commonly based on either finite difference methods \citep{malik1993lagrangian} or directly taking temporal derivatives of trajectory functions \citep{luthi2002some,gesemann2016noisy,schanz2016shake}. 
For instance, the first-order finite difference method \citep{malik1993lagrangian} evaluates a particle's velocity via dividing the particle displacement between two consecutive frames by the time interval.
The finite difference methods rely on the assumption that a particle travels a near-straight pathline within a short time interval.
However, when this assumption fails, it may lead to inaccurate reconstruction.
On the other hand, a trajectory-based velocity reconstruction first approximates particle pathlines by a continuous function, and next, velocities are derived from the definition of the velocity: that is the rate of change of a particle's location with respect to time.
However, similar to polynomial-based trajectory reconstruction, trajectory-based velocity reconstruction suffers from a lack of global smoothness, and the dilemma of selecting an appropriate order. 

Methods for reconstructing differential quantities can be classified into meshless and mesh-based approaches. 
In mesh-based methods (e.g., Flowfit by \citet{gesemann2016noisy}, Vortex-in-Cell+  by \citet{schneiders2016dense}, and Vortex-in-Cell\# by \citet{jeon2022fine}), Lagrangian data are first converted to an Eulerian mesh and differential quantities are then computed on the mesh using finite difference methods or some data conversion functions. 
However, relying on data conversions not only deviates from the original purpose of LPT but could introduce additional computational errors. 
LPT data typically exhibit irregular particle distributions (e.g., the distances between the particles may significantly vary 
 over the domain, in space and time), but the Eulerian meshes are often structured.
This mismatch between the Lagrangian data and Eulerian mesh is a classic challenge in the approximation theory \citep{fasshauer2007meshfree}. 
In addition to the data conversion errors, mesh-based methods cannot directly evaluate differential quantities at the location of Lagrangian data.
A typical solution is computing differential quantities at Eulerian meshes first and then interpolating Eulerian data back to the Lagrangian data points.

On the other hand, meshless methods \citep{luthi2002some,takehara2017direct,sperotto2022meshless,li2023three} 
avoid projecting Lagrangian data onto Eulerian mesh and can directly reconstruct (differential) flow quantities at scattered LPT data points.
These methods approximate velocity fields in each frame using continuous model functions, such as polynomials \citep{luthi2002some,takehara2017direct} and Radial Basis Functions (RBFs) \citep{sperotto2022meshless,li2023three}.
Subsequently, the differential quantities are calculated based on the spatial derivatives of the velocity field.
This approach is analogous to velocity reconstruction by taking the time derivative of the pathline.

The current mesh-based and meshless methods may face some difficulties. 
First, physical constraints may be absent in the reconstruction. 
The constraints can arise from some \textit{a priori} knowledge about the fluid flow (e.g.,~velocity solenoidal conditions for incompressible flows and boundary conditions \citep{gesemann2016noisy,jeon2022fine,sperotto2022meshless,li2023three}).
Neglecting these physical constraints could produce unrealistic or inaccurate flow structures in the reconstruction.
Second, processing large-scale LPT data can be computationally demanding and unstable. 
Consequently, the computational complexity grows rapidly as the number of particles and frames increases, potentially straining computational resources and prolonging processing time.
Some meshless methods, such as RBFs, may suffer from numerical instability when addressing a large number of data points without care being taken \citep{fornberg2004stable,fornberg2011stable}.
Third, the velocity field in each frame is sometimes described by low-order polynomials or B-splines.
While these functions provide the simplicity of computation, they may not be suitable for accurately capturing highly nonlinear flows.
Last but not least, velocities calculated by temporal derivatives of trajectories (in the Lagrangian perspective) are often decoupled from the velocity field reconstructed for each frame (in the Eulerian perspective). 
The former is derived from the definition of velocity, while the latter often respects some physical constraints (e.g., velocity divergence-free conditions). 
However, the velocity field of a given flow is supposed to be unique, no matter looking at it from the Eulerian or Lagrangian perspective.
Assimilating these velocities could improve reconstruction quality.

Several comprehensive flow field reconstruction strategies have been proposed. 
Typically, these strategies begin by initializing particle trajectories, velocities, and acceleration using some simple schemes, then these initialized quantities are used for downstream analysis or data assimilation. 
One such strategy, introduced by \citet{gesemann2016noisy}, comprises two algorithms: Trackfit and Flowfit.
The Trackfit initializes particle trajectories, velocities, and accelerations along particle pathlines based on the raw LPT data. 
Particle trajectories are approximated using cubic B-splines with time being its independent variable.
The velocity and acceleration are calculated by taking temporal derivatives of the trajectory functions.
On the other hand, the Flowfit algorithm is a data assimilation program.
It leverages the particle velocities and accelerations obtained from the Trackfit as inputs. 
In each frame, Flowfit converts the Lagrangian data onto an Eulerian mesh using 3D weighted cubic B-splines. 
Flowfit assimilates data by minimizing cost functions containing some constraints, such as divergence-free conditions for incompressible flows and curl-free conditions for pressure gradients. 
These constraints regularize the reconstructed flow quantities.

However, in Trackfit, the trajectory functions are embedded in a cost function that relies on the assumption of minimum change in acceleration.
While this practice is helpful to smooth the trajectories, it does not strictly reinforce physical conditions.
Additionally, to comply with this assumption, the time interval between adjacent frames must be small, which may not be suitable for data without time-resolved properties.
Last, converting Lagrangian data onto Eulerian meshes is needed in Flowfit. 

\citet{luthi2002some} proposed a meshless comprehensive reconstruction method.
This method first uses a localized cubic polynomial function to approximate particle trajectories along pathlines.
Second, velocities and accelerations are calculated by taking temporal derivatives of the trajectory functions.
Last, linear piece-wise polynomial functions are employed to approximate the velocity field in each frame, using velocities obtained from the previous step as inputs.
This method can be considered a more sophisticated version of Trackfit.
However, without further data assimilation, it cannot strictly enforce the velocity divergence-free constraint as intended.
Instead, it attempts to incorporate the divergence-free property by using a nonzero divergence as a normalized weight for weighted least squares to improve the reconstruction quality. 
Additionally, the use of linear piece-wise polynomial functions only provides piece-wise linear smoothness throughout the domain,
which may be inadequate to approximate complex velocity fields.

In the current research, we propose a meshless comprehensive reconstruction strategy for processing raw data measured by 3D TR-LPT systems.
This strategy incorporates (i) a stable RBF method \citep{fornberg2008stable,fornberg2011stable,larsson2013stable} to approximate particle trajectories and velocities along pathlines, as well as velocity fields and differential quantities in each frame;
(ii) the Constrained Least Squares (CLS) algorithm to enforce physical constraints and suppress noise; 
and (iii) the Partition-of-Unity Method (PUM)~\citep{melenk1996partition,babuvska1997partition} to reduce computational costs and improve numerical stability.
We refer to our strategy as the CLS-RBF PUM method.

This paper is organized as follows:
in Sect.~\ref{sec: mathematics tool}, the three foundation algorithms (i.e., the stable RBF, CLS, and PUM) are introduced.
Sect.~\ref{sec: cls-rbf pum method} elaborates on how these three foundation algorithms are integrated into one comprehensive reconstruction method.
Sect.~\ref{sec: results & discussion} shows validations of our method based on synthetic and experimental data.
Sect.~\ref{sec: conclusion} concludes the paper.

\section{Foundation Algorithms}
\label{sec: mathematics tool}
\subsection{Stable radial basis function (RBF)}
The classic RBF, also known as RBF-Direct \citep{fornberg2011stable}, is a kernel-based meshless algorithm for data approximation.
It uses the Euclidean norm between two data points as an independent variable.
With a specific kernel, such as a Gaussian  or multiquadric kernel, the RBF-Direct enjoys infinitely smoothness and can be easily extended to high dimensions.
Various versions of RBFs have been widely applied in computer graphics \citep{carr2001reconstruction,macedo2011hermite,drake2022implicit}, machine learning \citep{keerthi2003asymptotic,huang2006universal}, as well as flow field reconstruction in some recent works \citep{sperotto2022meshless,li2023three}.

However, as pointed out by \citet{fornberg2004stable}, the RBF-Direct faces a critical dilemma regarding its shape factor.
The shape factor controls the profile of RBF kernels:
a small shape factor, corresponding to the near-flat kernel, offers an accurate approximation but leads to an ill-conditioned problem.
On the other hand, a large shape factor, corresponding to a spiky kernel, provides a well-conditioned but inaccurate approximation.
It was conventionally believed that a trade-off must be made regarding the shape factor to strike a balance between accuracy and stability until stable RBFs emerged.

The stable RBFs can achieve numerical stability without compromising accuracy.
The ill-conditioning problem due to a small shape factor can be overcome by a handful of stable RBFs (e.g., RBF-CP by \citet{fornberg2004stable}, RBF-GA by \citet{fornberg2013stable}, RBF-RA by \citet{wright2017stable} and polynomial embedded RBF \citet{sperotto2022meshless}).
The RBF-QR \citep{fornberg2008stable,fornberg2011stable} is one such stable RBF.
The RBF-QR kernel $\psi$ is converted from an RBF-Direct kernel $\phi$, via a process of factorization, Taylor expansion, coordinate conversion, Chebyshev polynomial substitution, QR factorization, etc. 
The RBF-QR kernels enjoy well-conditioning and stability for any small shape factor, i.e., $\varepsilon \rightarrow 0^+$.
More details about the RBF-QR can be found in \citet{fornberg2011stable,larsson2013stable}, and here we briefly summarize its application for interpolation as an example. 

For an RBF-QR interpolation problem, given scalar data $\hat{f}_i^\text{c} \in \mathbb{R}$ located at a center $\hat{\bm{\xi}}_i^\text{c} \in \mathbb{R}^d$, i.e., $(\hat{\bm{\xi}}_i^\text{c},\hat{f}_i^\text{c})$, an interpolant $\tilde{s}(\varepsilon,\bm{\xi})$ can be written as a linear combination of $N$ RBF-QR kernels $\psi$:
\begin{equation*}
\tilde{s}(\varepsilon,\bm{\xi})=\sum_{i=1}^{N}\lambda_i\psi(\varepsilon, \Vert \bm{\xi}-\hat{\bm{\xi}}_i^\text{c}  \Vert )={\bf{\Psi}}(\varepsilon,\bm{\xi})\bm{\lambda},
\label{rbfqr}
\end{equation*}
where $N$~is the number of centers, $\bm{\lambda}=(\lambda_1,\lambda_2,\dots,\lambda_N)^{\text{T}}$~is the vector of expansion coefficients in which $\lambda_i$ controls the weights of the kernels, $\varepsilon$~is the shape factor, $\| \bm{\xi}-\hat{\bm{\xi}}_i^\text{c}\|$ denotes the Euclidean norm between evaluation points $\bm{\xi}$ and the center $\hat{\bm{\xi}}_i^\text{c}$.
The evaluation points $\bm{\xi}$ are where data are interpolated at.
The expansion coefficient $\lambda_i$ can be calculated by forcing the evaluation points coinciding with the centers, and then substituting interpolants with the given data: $\left. \tilde{s}(\varepsilon,\bm{\xi}) \right |_{\hat{\bm{\xi}}^{\text{c}}_i}=\tilde{s}(\varepsilon,\hat{\bm{\xi}}^{\text{c}}_i) =\hat{\bm{f}}^{\text{c}}$, where $\hat{\bm{f}}^{\text{c}}=(\hat{f}_1^\text{c},\hat{f}_2^\text{c},\dots,\hat{f}_N^\text{c})^{\text{T}}$ is the vector of the given data.

The derivative approximation can be calculated using the same expansion coefficients but with an RBF-QR derivative kernel $\psi_\mathcal{D}$:
\begin{equation*}
\tilde{s}_{\mathcal{D}}(\varepsilon,\bm{\xi})=\sum_{i=1}^{N}\lambda_i\psi_\mathcal{D} (\varepsilon,  \Vert \bm{\xi}-\hat{\bm{\xi}}_i^\text{c} \Vert )={\bf{\Psi}}_\mathcal{D}(\varepsilon,\bm{\xi})\bm{\lambda},
\label{rbfqr-deriv}
\end{equation*}
where $\mathcal{D}$ denotes a linear derivative operation.
With its scattered data approximation ability and easy derivative calculation, the RBF-QR is suitable for LPT data processing. 
In the CLS-RBF PUM method, we use the RBF-QR as a model function to approximate particle trajectories, velocities, and differential quantities. 
One can consider RBF-QR as a one-layer neural network that is fully `transparent' and interpretable, and the need for hyper-parameter tuning is minimized.

\subsection{Constrained least squares (CLS)}
\label{section: cls}
The CLS is based on the least squares regression with constraints enforced by the Lagrangian multiplier method.
The Lagrangian objective function $\mathcal{L}$ is created by appending an equality constraint ${\bf{C}} {\bm{\lambda}}=\bm{d}$ to a residual $\mathcal{R}$: 
\begin{equation}
    \mathcal{L}(\bm{\lambda},\bm{\eta})=\mathcal{R}+\bm{\eta}({\bf{C}} {\bm{\lambda}}-\bm{d}),
    \label{eq: lof}
\end{equation}
where $\mathcal{R}=\sum^N_i \Vert \tilde{s}(\varepsilon,\hat{\bm{\xi}}_{i}^{\text{c}})-\hat{f}_{i}^{\text{c}} \Vert^2$ is the residual between the measurements $\hat{f}_{i}^{\text{c}}$ and RBF-QR model function $\tilde{s}(\varepsilon,\hat{\bm{\xi}}_{i}^{\text{c}})={\bf{B}}(\varepsilon,\hat{\bm{\xi}}_{i}^{\text{c}})\bm{\lambda}$, where $ {\bf{B}}(\varepsilon,\hat{\bm{\xi}}_{i}^{\text{c}})=B_{ij}=\psi (\varepsilon,\Vert \hat{\bm{\xi}}_i^{\text{c}}- \bm{\xi}_j^{\text{ref}} \Vert ) $
is the RBF-QR system matrix constructed by $N$ centers $\hat{\bm{\xi}}_{i}^{\text{c}}$ and $M$ reference points $\bm{\xi}_{j}^{\text{ref}}$, $i=1,2,\dots,N$ and $j=1,2,\dots,M$.
$\bm{\eta} = (\eta_1,\eta_2,\dots,\eta_J)^{\text{T}}$ is the vector of Lagrangian multipliers.
$\bf{C}$ is a generalized constraint matrix;
it can be a constraint matrix ${\bf{C}}_\mathcal{O}$ for function values and/or ${\bf{C}}_\mathcal{D}$ for function derivatives.
$\bf{C}$ is established by $J$ constraint points $\bm{\xi}_{l}^{\text{cst}}$ and $M$ reference points $\bm{\xi}_{j}^{\text{ref}}$ with entries
\begin{equation*}
\begin{split}
    {\bf{C}}_\mathcal{O}(\varepsilon,\bm{\xi}_{l}^{\text{cst}})=C_{\mathcal{O},lj}&=\psi(\varepsilon,\Vert {\bm{\xi}}_l^{\text{cst}}- \bm{\xi}_j^{\text{ref}}\Vert ) \\
    {\bf{C}}_\mathcal{D}(\varepsilon,\bm{\xi}_{l}^{\text{cst}})=C_{\mathcal{D},lj}&=\psi_{\mathcal{D}}(\varepsilon,\Vert {\bm{\xi}}_l^{\text{cst}}- \bm{\xi}_j^{\text{ref}}\Vert )
\end{split},
\end{equation*}
$l=1,2,\dots,J$.
$\bm{\lambda} = (\lambda_1,\lambda_2,\dots,\lambda_M)^{\text{T}}$ is the vector of expansion coefficients.
$\bm{d}$ is the vector of constraint values;
in this work, $\bm{d}$ is a null vector to comply with the divergence-free constraint.
An oversampling ratio is defined as $\beta=N/M$.
$\beta>1$ is required for regression and a sufficiently large $\beta$ provides smooth reconstruction.

Next, we minimize the Lagrangian objective function Eq.~\eqref{eq: lof} for expansion coefficients that will be used for approximation.
By setting the gradient of $\mathcal{L}$ with respect to the vectors $\bm{\lambda}$ and $\bm{\eta}$ to zero (i.e.,  $\partial{\mathcal{L}}/\partial{\bm{\lambda}} =0$ and $\partial{\mathcal{L}}/\partial{\bm{\eta}} =0$), a linear system is established:
\begin{equation}
\begin{bmatrix}
    {\bf{G}} & {\bf{C}}^{\text{T}}\\
    \bf{C} & \bf{0}
\end{bmatrix}
\begin{bmatrix}
    {\bm{\lambda}} \\ {\bm{\eta}}
\end{bmatrix}=
\begin{bmatrix}
   {\bf{F}} \\ \bm{d}
\end{bmatrix},
\label{eqn: cls_rbfqr}
\end{equation}
where ${\bf{G}}={\bf{B}}^{\text{T}}(\varepsilon,\hat{\bm{\xi}}_{i}^{\text{c}}){\bf{B}}(\varepsilon,\hat{\bm{\xi}}_{i}^{\text{c}})$ and ${\bf{F}}={\bf{B}}^{\text{T}}(\varepsilon,\hat{\bm{\xi}}_{i}^{\text{c}}) \hat{\bm{f}}^{\text{c}}$. 
After solving $\bm{\lambda}$ 
from Eq.~\eqref{eqn: cls_rbfqr}, the RBF-QR approximation and its derivative functions are calculated by:
\begin{equation}
\begin{split}
    \tilde{s}(\varepsilon,\bm{\xi}_k^{\text{eva}}) &={\bf{E}}(\varepsilon,\bm{\xi}_k^{\text{eva}})\bm{\lambda} \\
    \tilde{s}_{\mathcal{D}}(\varepsilon,\bm{\xi}_k^{\text{eva}}) &={\bf{E}}_{\mathcal{D}}(\varepsilon,\bm{\xi}_k^{\text{eva}})\bm{\lambda}
\end{split},
\label{eqn: lsrbfqr_evaluation}
\end{equation}
where the RBF-QR evaluation matrix ${\bf{E}}(\varepsilon,\bm{\xi}_{k}^{\text{eva}})$ and its derivative matrix ${\bf{E}}_{\mathcal{D}}(\varepsilon,\bm{\xi}_{k}^{\text{eva}})$ are constructed by $P$ evaluation points $\bm{\xi}^{\text{eva}}_k$ and $M$ reference points $\bm{\xi}^{\text{ref}}_j$ with entries:
\begin{equation}
\begin{split}
        {\bf{E}}(\varepsilon,\bm{\xi}_{k}^{\text{eva}}) &=E_{kj}=\psi (\varepsilon,\Vert \bm{\xi}^{\text{eva}}_k- \bm{\xi}_j^{\text{ref}}\Vert ) \\
        {\bf{E}}_{\mathcal{D}}(\varepsilon,\bm{\xi}_{k}^{\text{eva}}) &=E_{\mathcal{D},kj}=\psi_{\mathcal{D}} (\varepsilon,\Vert \bm{\xi}_k^{\text{eva}}- \bm{\xi}_j^{\text{ref}}\Vert )
\end{split},
\label{eqn: lsrbfqr_evaluationmatrix}
\end{equation} 
where $k=1,2,\dots,P$. 

Eqs.~\eqref{eqn: cls_rbfqr} and \eqref{eqn: lsrbfqr_evaluation} cannot be directly applied in 3D while subject to divergence-free constraints without proper extensions.
The matrix elements in these two equations must be extended to each direction of the coordinates since the reconstruction is in 3D and divergence-free constraints consist of derivatives in three directions.
The extended linear system is written as
\begin{equation}
\begin{bmatrix}
    \bar{\bf{G}} & \bar{\bf{C}}^{\text{T}}\\
    \bar{\bf{C}} & \bf{0}
\end{bmatrix}
\begin{bmatrix}
    \bar{\bm{\lambda}} \\ \bar{\bm{\eta}}
\end{bmatrix}=
\begin{bmatrix}
   \bar{\bf{F}} \\ \bar{\bm{d}}
\end{bmatrix}.
\label{eqn: cls_rbfqr_ex}
\end{equation}
In Eq.~\eqref{eqn: cls_rbfqr_ex}, 
 $\bar{\bf{C}}=\begin{bmatrix}{{\bf{C}}}_x & {\bf{C}}_y & {\bf{C}}_z\end{bmatrix}
    \label{extend_constraint}$ is the extended constraint matrix, 
where ${\bf{C}}_x$, ${\bf{C}}_y$, and ${\bf{C}}_z$ are first-order spatial derivative constraint matrices based on ${\bf{C}}_\mathcal{D}$ in the $x$, $y$, and $z$ directions, respectively.
The extended matrices $\bar{\bf{G}}$ and $\bar{\bf{F}}$ are block diagonal matrices with entries
\begin{equation*}
\begin{split}
    \bar{\bf{G}}=
    \begin{bmatrix}
    {\bf{B}}^{\text{T}}{\bf{B}} & & \\
    & {\bf{B}}^{\text{T}}{\bf{B}} & \\
    & & {\bf{B}}^{\text{T}}{\bf{B}}
    \end{bmatrix},~
    \bar{\bf{F}} =
    \begin{bmatrix}
    {\bf{B}}^{\text{T}} & & \\
    & {\bf{B}}^{\text{T}} & \\
    & & {\bf{B}}^{\text{T}}
    \end{bmatrix}\hat{\bm{f}}^{\text{c}},
\end{split}
\label{extend_G}
\end{equation*}
where $\hat{\bm{f}}^{\text{c}}=\begin{pmatrix} \bm{u} & \bm{v} & \bm{w} \end{pmatrix}^{\text{T}}$, and $\bm{u}$, $\bm{v}$, and $\bm{w}$ are the velocity vectors in the $x$, $y$, and $z$ directions, respectively;
$\bar{\bm{d}}$ is a null column vector corresponding to the divergence-free constraints.
After solving $ \bar{\bm{\lambda}}$ in Eq.~\eqref{eqn: cls_rbfqr_ex},
the CLS RBF-QR approximation function $\tilde{s}(\varepsilon,\bm{\xi}_k^{\text{eva}})$ and its differentiation function $\tilde{s}_{\mathcal{D}}(\varepsilon,\bm{\xi}_k^{\text{eva}})$ are calculated by:
\begin{equation}
\begin{split}
    \tilde{s}(\varepsilon,\bm{\xi}_k^{\text{eva}}) &=\bar{\bf{E}}(\varepsilon,\bm{\xi}_k^{\text{eva}})\bar{\bm{\lambda}} \\
    \tilde{s}_{\mathcal{D}}(\varepsilon,\bm{\xi}_k^{\text{eva}}) &=\bar{\bf{E}}_{\mathcal{D}}(\varepsilon,\bm{\xi}_k^{\text{eva}})\bar{\bm{\lambda}}
\end{split},
\label{cls_rbfqr_eval}
\end{equation}
where $\bar{\bf{E}}$ and $\bar{\bf{E}}_{\mathcal{D}}$ are extended diagonal block matrices based on ${\bf{E}}$ and ${\bf{E}}_{\mathcal{D}}$ in Eq.~\eqref{eqn: lsrbfqr_evaluationmatrix}, respectively:
\begin{equation*}
\begin{split}
    \bar{\bf{E}} =
    \begin{bmatrix}
    {\bf{E}} & & \\
    & {\bf{E}} & \\
    & & {\bf{E}}
    \end{bmatrix},~
    \bar{\bf{E}}_{\mathcal{D}} =
    \begin{bmatrix}
    {\bf{E}}_{\mathcal{D}} & & \\
    & {\bf{E}}_{\mathcal{D}} & \\
    & & {\bf{E}}_{\mathcal{D}}
    \end{bmatrix}
\end{split}.
\end{equation*}
Up to this point, a CLS RBF-QR framework is established for a 3D Lagrangian flow field reconstruction with divergence-free constraints as an example. 
This method can enforce other constraints in a similar fashion, if needed.

\begin{figure}[!htb]
	\centering
	\includegraphics[scale=0.13]{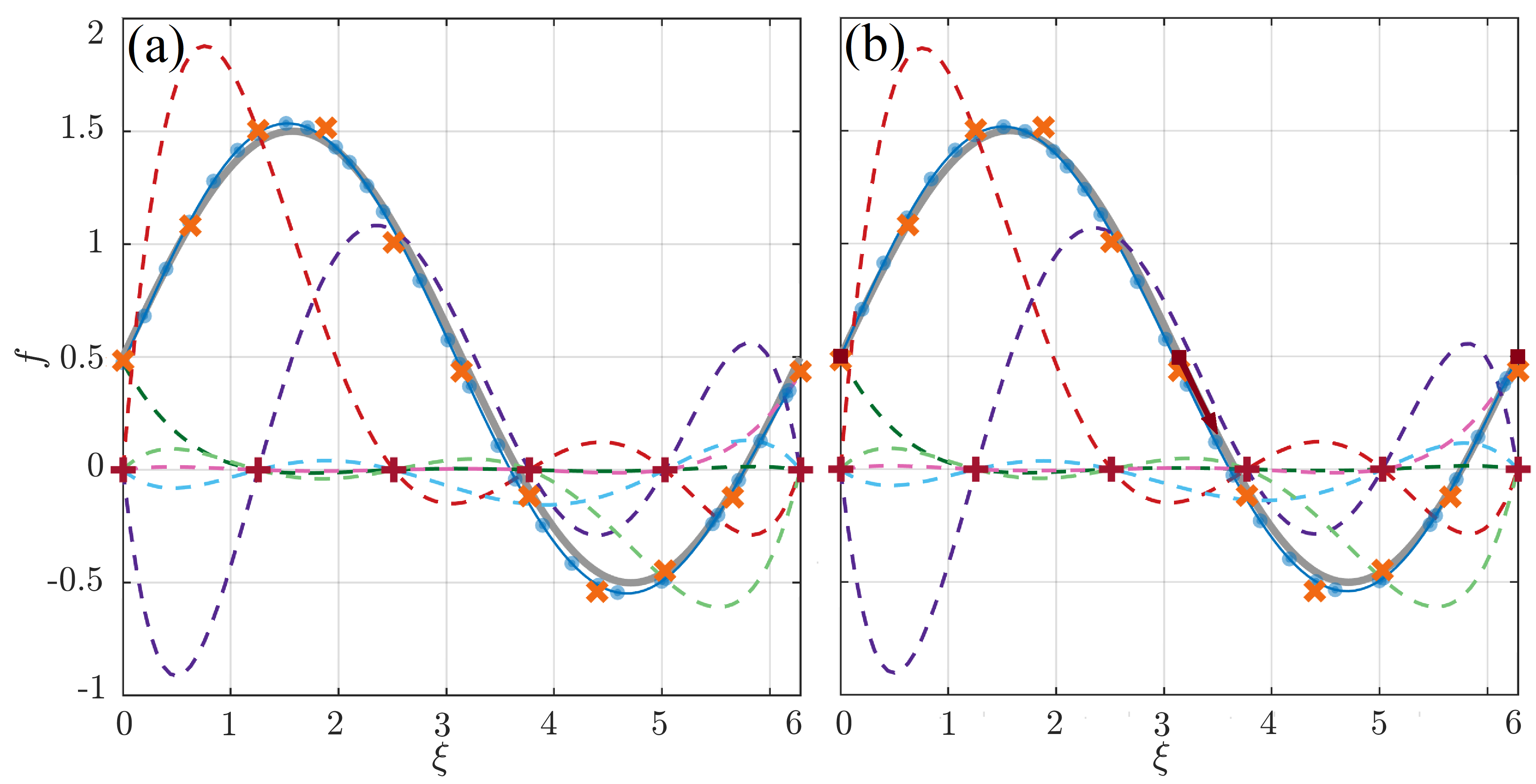}
	\caption{A demonstration of RBF-QR approximation and centers, reference points, constraint points, and evaluation points in 1D. 
 The RBF-QR (a) unconstrained and (b) constrained regression have six kernels. Gray solid curves: the `ground truth' based on an exact function $f(\xi)=\sin{(\xi)}+0.5$, $\xi \in [0,2\pi]$; blue solid curves: RBF-QR reconstruction; dashed curves: RBF-QR kernels `centered' at different reference points; orange crosses: the given data, which are sampled from the ground truth with random perturbation to simulate the corrupted experimental data; scarlet crosses: the reference points; scarlet squares: constraints of function values or function derivatives; the scarlet square with an arrowhead indicates the derivative constraint; blue dots: reconstructed results at the evaluation points.}
\label{fig:ep_ls}
\end{figure}

The CLS-RBF PUM method relies on four types of points (listed below), each playing a distinct role in the flow reconstruction. 
We use one-dimensional (1D) unconstrained and constrained RBF-QR regression for demonstration as shown in Fig.~\ref{fig:ep_ls}. 
\begin{enumerate}
\item Centers $\hat{\bm{\xi}}^\text{c}$ ($x$ coordinates of the orange crosses): they are locations of the given data.
The centers are determined by experiments and are typically `randomly' scattered throughout the flow domain.

\item Reference points $\bm{\xi}^\text{ref}$ ($x$ coordinates of scarlet crosses): a linear combination of kernels (see dashed curves) centered at the reference points can approximate the given data with a continuous function (i.e., the approximation function $\tilde{s}(\bm{\xi})$, indicated by the blue solid curves).
Note, the number of reference points should be smaller than that of the centers to ensure effective regression.
Besides, the placement of the reference points prefers a quasi-uniform layout such as the Halton points for 2D and 3D, and a uniform point layout for 1D \citep{fasshauer2007meshfree, Larsson2023privaite}.

\item Constraint points ${\bm{\xi}}^{\text{cst}}$ ($x$ coordinates of scarlet squares): they are where physical constraints are enforced.
In the current work, we impose divergence-free constraints at the centers to guarantee that velocity divergence at measured LPT data points is zero.
Generally, there is no limitation to the placement of the constrained points (e.g., locations or numbers of the constraints), as long as the linear system in Eqs.~\eqref{eqn: cls_rbfqr} and \eqref{eqn: cls_rbfqr_ex} is well-posed. 

\item Evaluation points $\bm{\xi}^\text{eva}$ ($x$ coordinates of blue dots): the locations where $\tilde{s}(\bm{\xi})$ is reconstructed are the evaluation points. 
The number or the locations of evaluation points have no limitations. 
This means that the evaluation points can be densely placed in the domain to achieve super-resolution or placed at the centers $\hat{\bm{\xi}}^\text{c}$ to directly evaluate at locations of LPT data.
\end{enumerate}

\subsection{Partition-of-unity method (PUM)}
\label{subsection: pum}
The PUM \citep{melenk1996partition,babuvska1997partition} is applied in reconstruction to further enhance numerical stability and efficiency.
The ill-conditioning issue in the RBF-Direct is not only caused by a small shape factor but also by a large-scale data set \citep{fornberg2004stable,fornberg2011stable}.
The same ill-conditioning problem due to large-scale data persists in stable RBFs.
In addition, processing a large number of data at once can be prohibitively expensive. 

\begin{figure}[ht]
	\centering
	\includegraphics[scale=0.12]{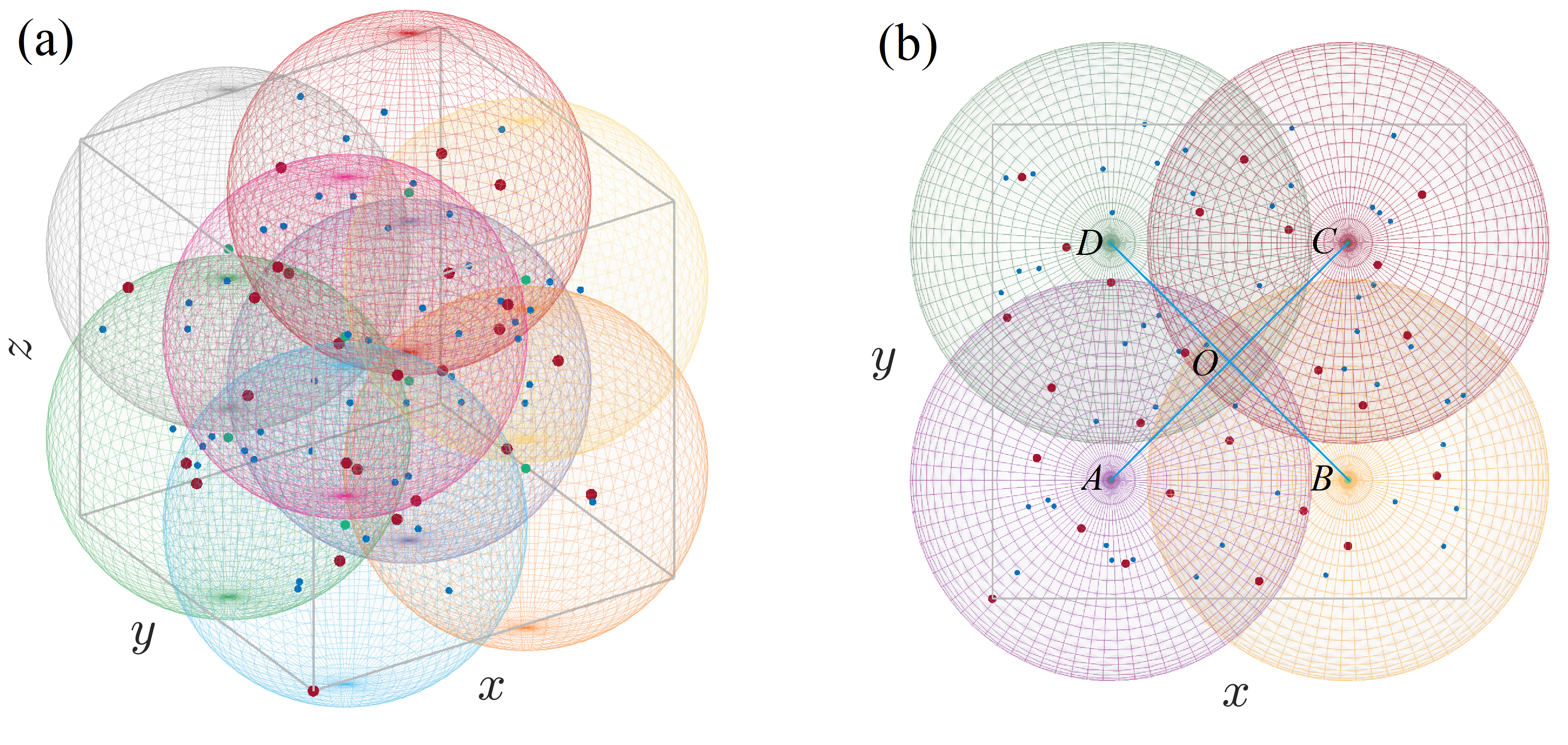}
	\caption[A demonstration of least squares RBF-QR fitting using centers and reference points in one dimension]
	{Example PUM patch layout in 3D. The domain is partitioned by eight spherical PUM patches. (b) is the top view of (a). The gray lines indicate the edges of the domain $\Omega$. Spheres in different colors are PUM patches, whose centers are marked by green dots; red dots indicate the reference points of a Halton layout; randomly distributed blue dots are the given data points. 
 }
\label{fig:pum}
\end{figure}

The PUM can partition the domain and localize flow field reconstruction.
Here, we briefly summarize the implementation of RBF-QR with PUM, which was introduced by \citet{larsson2017least}.
First, identical spherical PUM patches $\Omega_m$, $m=1, 2, \dots, N_P$, are created to cover the entire 3D flow domain $\Omega$.
$N_P$ is the number of PUM patches.
These patches are overlapped and the overlap ratio is defined as $\gamma$. 
The radius of PUM patches is calculated as $\rho=(1+\gamma)\rho_0$, where $\rho_0$ is the radius of patches that have no overlaps in the diagonal direction.
For example in Fig.~\ref{fig:pum}b, $\rho_0 =\overline{AO}$, and the diagonal directions refer to the lines $\overline{AC}$ and $\overline{BD}$.
Next, every patch is assigned a weight function $W_{m}(\bm{\xi})$. 
The weight function becomes zero outside a patch, i.e., $W_{m}(\bm{\xi})=0$;
and the sum of all weight functions from all patches at an arbitrary point in the domain is unity, i.e., ${\textstyle\sum_{m=1}^{N_P}{W_{m}(\bm{\xi})}=1}$. 
The weight functions are based on Shepard’s method \citep{shepard1968two} following Larsson's work: 
\begin{equation*}
    W_{m}(\bm{\xi})=\frac{\varphi_{m}(\bm{\xi})}{{{\textstyle{\sum_{m=1}^{N_P}\varphi_{m}(\bm{\xi})}}}},
\end{equation*}
where $\varphi_m(\bm{\xi})$ is a compactly supported generating function, and the Wendland $C^2$ function $\varphi(r)=(1-r^4)_{+}(4r+1)$ \citep{wendland1995piecewise} is chosen here. 
Last, the global evaluation function on the fluid domain $\Omega$ can then be assembled by a weighted summation of local approximation functions: 
\begin{equation}
\begin{split}
    \tilde{s}(\varepsilon,\bm{\xi})&=\sum_m^{N_p} W_{m}(\bm{\xi})\tilde{s}_{m}(\varepsilon,\bm{\xi}) \\ 
    \tilde{s}_{\mathcal{D}}(\varepsilon,\bm{\xi})&=\sum_m^{N_p} W_{\mathcal{D},m}(\bm{\xi})\tilde{s}_{m}(\varepsilon,\bm{\xi})+ W_{m}(\bm{\xi})\tilde{s}_{\mathcal{D},m}(\varepsilon,\bm{\xi})
\end{split},
\label{pum}
\end{equation}
where $W_{\mathcal{D},m}(\bm{\xi})$ is linear derivatives of weight functions in the patch $m$ and $\tilde{s}_{\mathcal{D},m}(\varepsilon,\bm{\xi})$ is a linear derivative approximation in the same patch, both are derived by the chain rule.
Fig.~\ref{fig:pum} presents a 3D example of PUM patches in a unit cubic domain with an overlap ratio $\gamma=0.2$.

\section{CLS-RBF PUM Method} 
\label{sec: cls-rbf pum method}
\begin{figure}[ht]
    \centering
    \includegraphics[scale=0.27]{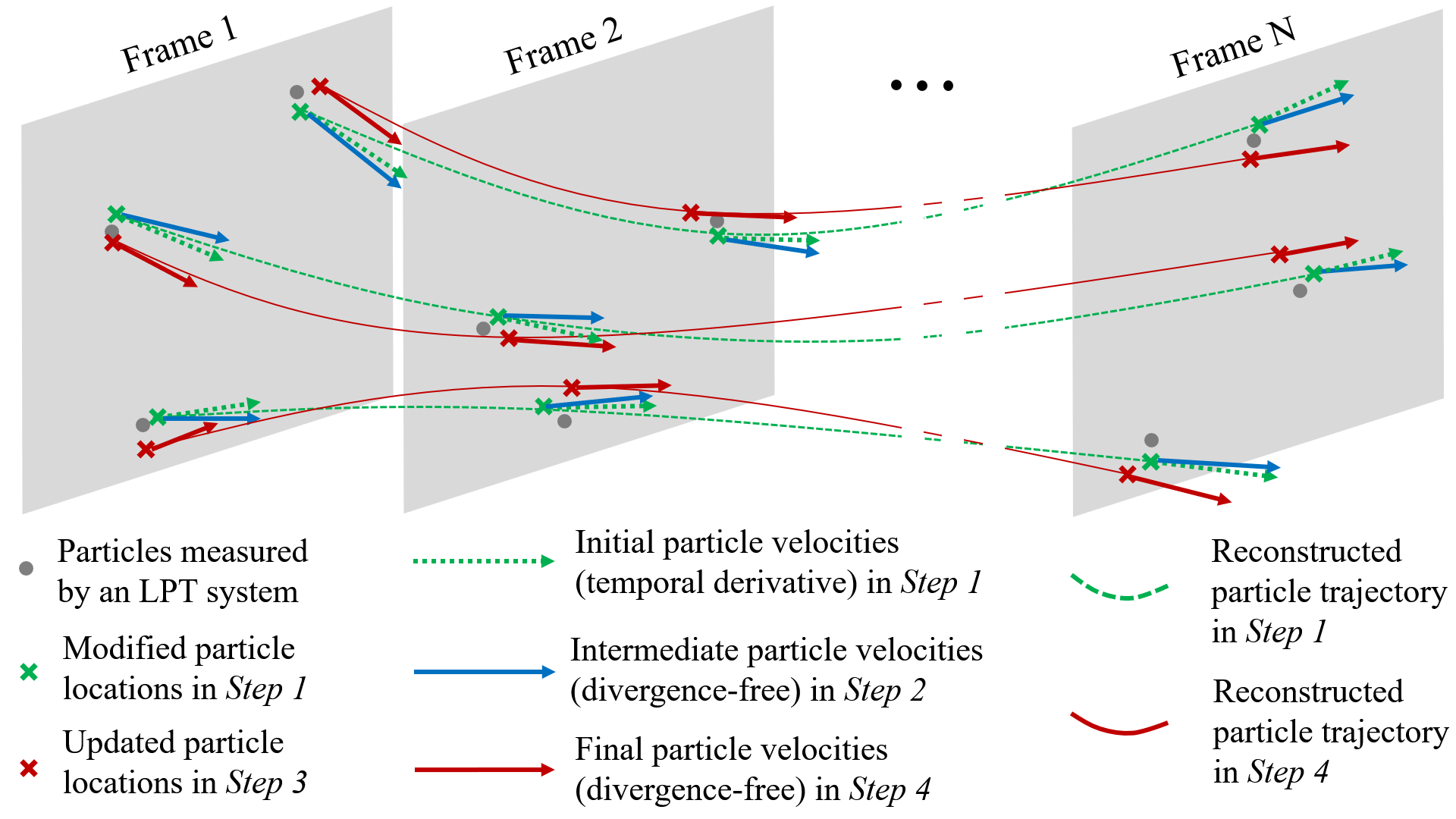}
    \caption{A 2D demonstration of CLS-RBF PUM method. Only three out of $N$ frames and reconstruction at three particles are highlighted here.}
    \label{fig:2D_demo}
\end{figure}

We integrate the aforementioned three foundation algorithms (i.e., BRF-QR, CLS, and PUM) as one comprehensive reconstruction method. 
This method consists of four steps, which will be elaborated on in this section. 
An important byproduct of this method is the capability of super-resolution in time and space, which will be discussed lastly in this section. 
Hereafter, the generalized independent variable $\bm{\xi}$ used in Sect.~\ref{sec: mathematics tool} is substituted by particle spatial coordinates $\bm{x}$ or time $t$.
Fig.~\ref{fig:2D_demo} sketches the CLS-RBF PUM method processing TR-LPT data in 2D as a demonstration.

\subsection{Step 1: initialize particle trajectory and velocity} 
\label{section: step 1}
\textit{Step~1} initializes smooth particle trajectories (and velocities) by fitting the particle spatial coordinates provided by an LPT system.
This fitting is based on Eq.~\eqref{eqn: cls_rbfqr} with time being its independent variable, without any constraints.

Trajectory fitting is performed for all coordinates of each particle. 
Here, we use the $x$ coordinate of a particle as an example.
The vector of spatial coordinates $\hat{\bm{x}}=(\hat{x}_1,\hat{x}_2,\dots,\hat{x}_{N_\text{trj}})^\text{T}$ are measured at $N_{trj}$ time instants $ \bm{t}^{\text{c}}= (t^{\text{c}}_{1},t^{\text{c}}_{2},\dots,t^{\text{c}}_{N_{\text{trj}}})^{\text{T}}$ by an LPT system. 
We refer to $\bm{t}^{\text{c}}$ as the vector of temporal centers.
Based on Eq.~\eqref{eqn: lsrbfqr_evaluation},
the trajectory function is given by:
\begin{equation}
    \tilde{\bm{x}}(\varepsilon,\bm{t}^{\text{c}})={\bf{E}}(\varepsilon,\bm{t}^{\text{c}})\bm{\lambda}^{\text{trj}}={\bf{E}}(\varepsilon,\bm{t}^{\text{c}}){\bf{B}}^{+}(\varepsilon,\bm{t}^{\text{c}})\hat{\bm{x}},
\label{traj_eval}
\end{equation}
where ${\bf{B}}^{+}(\varepsilon,\bm{t}^{\text{c}})=(({\bf{B}}^{\text{T}}(\varepsilon,\bm{t}^{\text{c}}){\bf{B}}(\varepsilon,\bm{t}^{\text{c}}))^{-1}{\bf{B}}^{\text{T}}(\varepsilon,\bm{t}^{\text{c}})$ is a generalized inverse of the RBF-QR system matrix ${\bf{B}}(\varepsilon,\bm{t}^{\text{c}})$ that has entries:
\begin{equation*}
     {\bf{B}}(\varepsilon,\bm{t}^{\text{c}})=B_{ij}=\psi (\varepsilon,\Vert \bm{t}_i^{\text{c}}- \bm{t}_j^{\text{ref}}\Vert ),
\end{equation*} 
where $i=1,2,\dots,N_{\text{trj}}$ and $j=1,2,\dots,M_{\text{trj}}$;
${\bm{t}}^{\text{ref}}=({t}^{\text{ref}}_{1},{t}^{\text{ref}}_{2},\dots,{t}^{\text{ref}}_{M_{\text{trj}}})^{\text{T}}$ is the vector of temporal reference points. 
The trajectory evaluation matrix ${\bf{E}}(\varepsilon,\bm{t}^{\text{c}})$ has the entries:
\begin{equation*}
    {\bf{E}}(\varepsilon,\bm{t}^{\text{c}})=E_{ij}=\psi (\varepsilon,\Vert \bm{t}^{\text{c}}_i- \bm{t}_j^{\text{ref}}\Vert ).
\label{step1_3}
\end{equation*}
A temporal oversampling ratio is defined as $\beta_0=N_{\text{trj}}/M_{\text{trj}}$, and
 $\beta_0>1$ is essential for regression.

The initial velocity is calculated based on the temporal derivatives of trajectory functions:
\begin{equation}
    \tilde{\bm{u}}(\varepsilon,\bm{t}^{\text{c}}) ={\bf{E}}_t (\varepsilon,\bm{t}^{\text{c}}) \bm{\lambda}^{\text{trj}}={\bf{E}}_t (\varepsilon,\bm{t}^{\text{c}}) {\bf{B}}^{+}(\varepsilon,\bm{t}^{\text{c}})\hat{\bm{x}}, 
\label{step1_velocity_acceleration}
\end{equation}
where  ${\bf{E}}_t (\varepsilon,\bm{t}^{\text{c}})$ is the velocity evaluation matrix based on Eq.~\eqref{eqn: lsrbfqr_evaluationmatrix} with $[\cdot]_t $ denoting the first-order temporal derivative:
\begin{equation*}
      {\bf{E}}_t (\varepsilon,\bm{t}^{\text{c}}) =  {\bf{E}}_t (\varepsilon,\bm{t}^{\text{c}}) =E_{t,ij}=\psi_{t} (\varepsilon,\Vert \bm{t}^{\text{c}}_i- \bm{t}_j^{\text{ref}}\Vert ).
\label{step1_5}
\end{equation*}
Acceleration can be computed accordingly if needed. 
Reconstruction of trajectories and velocities for $y$ and $z$ coordinates are similar. 
After computing the velocities in all three directions, the velocity field at each measured particle location in each frame is known in turn.

Hereafter, the particle location output from \textit{Step~1} is referred to as the modified particle location, as the raw particle locations are slightly modified by the regression process. 
The computed velocity fields from this step are called initial velocities as the velocity of each particle or the velocity field in each frame is known for the first time. 
Note, the initial velocities are computed from Lagrangian perspectives based on the definition of the velocity; however, they are not subject to any physical constraints.

\subsection{Step 2: calculate intermediate divergence-free velocity field} 
\label{section: step 2}
\textit{Step~2} calculates an intermediate divergence-free velocity field in each frame by constrained least squares.
This step uses the modified particle locations and initial velocities as the inputs.
To calculate the intermediate velocity fields, the matrices $\bf{B}$, $\bf{C}$, $\bf{E}$, and ${\bf{E}}_\mathcal{D}$  (described in Sect.~\ref{section: cls}) are constructed first.
For example in the $\kappa$-th frame, an RBF-QR spatial system matrix ${\bf{B}}(\varepsilon,\tilde{\bm{x}}^{\text{c}})$ is formulated based on $N$ modified spatial points $\tilde{\bm{x}}^{\text{c}}_i$ and $M_1$ spatial reference points $\bm{x}^{\text{ref}}_j$ with entries:
\begin{equation*}
    {\bf{B}}(\varepsilon,\tilde{\bm{x}}^{\text{c}})=B_{ij} =\psi (\varepsilon,\Vert \tilde{\bm{x}}^{\text{c}}_i - \bm{x}^{\text{ref}}_j\Vert ),
\end{equation*}
where
$\tilde{\bm{x}}^{\text{c}}=[\tilde{\bm{x}}^{\text{c}}_1,\tilde{\bm{x}}^{\text{c}}_2,\dots,\tilde{\bm{x}}^{\text{c}}_N]^\text{T}$ and $\tilde{\bm{x}}_i^{\text{c}}=(\tilde{x},\tilde{y},\tilde{z})^{\text{c}}_i$; 
$\bm{x}^{\text{ref}}=[\bm{x}^{\text{ref}}_1, \bm{x}^{\text{ref}}_2, \dots,\bm{x}^{\text{ref}}_{M_1}]^\text{T}$, and  $\bm{x}^{\text{ref}}_j=({x},{y},{z})^{\text{ref}}_j$, $i=1,2,\dots,N$, and $j=1,2,\dots,M_1$.
An RBF-QR spatial derivative constraint matrix ${\bf{C}}_{\mathcal{D}}(\varepsilon,\tilde{\bm{x}}^{\text{c}})$ is established between $N$ modified spatial centers $\tilde{\bm{x}}^{\text{c}}_i$ and $M_1$ spatial reference points $\bm{x}^{\text{ref}}_j$:
\begin{equation*}
    {\bf{C}}_{\mathcal{D}}(\varepsilon,\tilde{\bm{x}}^{\text{c}})=C_{\mathcal{D},ij} =\psi_{\mathcal{D}} (\varepsilon,\Vert \tilde{\bm{x}}^{\text{c}}_i- \bm{x}^{\text{ref}}_j\Vert ).
\label{step2_2}
\end{equation*}
Here, the constraint points $\bm{x}^{\text{cst}}$ coincide with modified particle locations $\tilde{\bm{x}}^{\text{c}}$ since we enforce the velocity divergence at the measured particle location to be zero.
The RBF-QR spatial evaluation matrix ${\bf{E}}(\varepsilon,\tilde{\bm{x}}^{\text{c}})$ is established by $N$ modified particle locations $\tilde{\bm{x}}^{\text{c}}_i$ and $M_1$ spatial reference points $\bm{x}^{\text{ref}}_j$:
\begin{equation*}
    {\bf{E}}(\varepsilon,\tilde{\bm{x}}^{\text{c}}) = E_{ij} =\psi (\varepsilon,\Vert \tilde{\bm{x}}^{\text{c}}_i- \bm{x}^{\text{ref}}_j\Vert ).
\label{step2_3}
\end{equation*}
In \textit{Step~2}, the evaluation points are placed at the same locations as the modified spatial centers.
A spatial oversampling ratio in {\it Step~2} is defined as $\beta_1=N/M_1$ and chosen to be slightly larger than unity.

After constructing the extended matrices, the expansion coefficients $\bar{\bm{\lambda}}$ is solved from Eq.~\eqref{eqn: cls_rbfqr_ex}, and the intermediate velocity  field $\tilde{\bf{U}}^{\text{div}}_{\kappa} =(\tilde{\bm{u}}^{\text{div}}_{\kappa},\tilde{\bm{v}}^{\text{div}}_{\kappa},\tilde{\bm{w}}^{\text{div}}_{\kappa})^{\text{T}}$ is computed based on Eq.~\eqref{cls_rbfqr_eval}:
\begin{equation}
    \tilde{{\bf{U}}}^{\text{div}}_{\kappa}(\varepsilon,\tilde{\bm{x}}^{\text{c}}) =\bar{\bf{E}}(\varepsilon,\tilde{\bm{x}}^{\text{c}})\bar{\bm{\lambda}}.
\label{cls_rbfqr_U_calc}
\end{equation}
The velocity fields $\tilde{{\bm{u}}}^{\text{div}}_{\kappa}$, $\tilde{{\bm{v}}}^{\text{div}}_{\kappa}$, and $\tilde{{\bm{w}}}^{\text{div}}_{\kappa}$ can be extracted from $\tilde{{\bf{U}}}^{\text{div}}_{\kappa}(\varepsilon,\tilde{\bm{x}}^{\text{c}})$.
The velocities reconstructed in \textit{Step~2} are divergence-free from the Eulerian perspective. 
However, they are not necessarily the same as the velocity fields obtained from \textit{Step~1}, which are based on the definition of the velocity but are not divergence-free. 
This discrepancy is due to errors in measured particle spatial coordinates. 
To resolve this conflict, we assimilate the results from \textit{Steps 1} and \textit{2} in \textit{Step 3}.

\subsection{Step 3: update particle location by data assimilation} 
\label{section: step 3}
\textit{Step~3} incorporates the Lagrangian and Eulerian reconstructions by updating the particle trajectories using least squares regression for all particles in all frames.
The underlying motivation is that the velocities calculated by temporal derivatives of trajectories for each particle (i.e., velocities output from \textit{Step~1}, which is a Lagrangian reconstruction) should be identical to the velocities reconstructed by the constrained regression in each frame (i.e., velocities output from \textit{Step~2}, which is an Eulerian reconstruction).
However, due to the errors in the measured particle spatial coordinates, these two velocity reconstructions are not necessarily equal to each other for the same flow field.
Nevertheless, the velocities calculated in \textit{Step~2} are assumed to be more accurate than those in \textit{Step~1} since they respect physical constraints (incompressibility condition in this case).
Therefore, solenoidal velocities output from \textit{Step 2} can be used to update particle locations.

The expansion coefficients of the trajectory function are re-calculated to update particle locations.
The update in particles' $x$ coordinates is presented as an example.
First, a linear system is constructed based on a modified trajectory matrix $\tilde{\bf{X}}(\varepsilon,\bm{t}^\text{c})$ and a divergence-free velocity matrix $\tilde{\bf{V}}(\varepsilon,\tilde{\bm{x}}^{\text{c}})$:
\begin{subequations}
\begin{align}
    \tilde{\bf{X}}(\varepsilon,\bm{t}^\text{c}) &={\boldsymbol{\Lambda}}{\bf{E}}^\text{T}(\varepsilon,\bm{t}^{\text{c}}),
    \label{step3_ls:a} \\
    \tilde{\bf{V}}(\varepsilon,\tilde{\bm{x}}^{\text{c}}) &={\boldsymbol{\Lambda}}{\bf{E}}_{t}^\text{T}(\varepsilon,\bm{t}^{\text{c}}),
    \label{step3_ls:b}
\end{align}
\end{subequations}
where ${\boldsymbol{\Lambda}}$ is an expansion coefficient matrix to be determined.
The elements in $\tilde{\bf{X}}(\varepsilon,\bm{t}^{\text{c}})$ is from Eq.~\eqref{traj_eval} in \textit{Step~1}, and $\tilde{\bf{X}}(\varepsilon,\bm{t}^{\text{c}})$ has entries:
\begin{equation*}
    \begin{split}
        \tilde{\bf{X}}(\varepsilon,\bm{t}^{\text{c}}) &= 
    \begin{bmatrix}
    \tilde{{x}}_{1}(t^{\text{c}}_1) & \tilde{{x}}_{1}(t^{\text{c}}_2) & \hdots & \tilde{{x}}_{1}(t^{\text{c}}_{N_{\text{trj}}}) \\
    \tilde{{x}}_{2}(t^{\text{c}}_1) & \tilde{{x}}_{2}(t^{\text{c}}_2) & \hdots & \tilde{{x}}_{2}(t^{\text{c}}_{N_{\text{trj}}}) \\
    \vdots & \vdots & \ddots & \vdots & \\
    \tilde{{x}}_{N}(t^{\text{c}}_1) & \tilde{{x}}_{N}(t^{\text{c}}_2) & \hdots & \tilde{{x}}_{N}(t^{\text{c}}_{N_{\text{trj}}}) \\
    \end{bmatrix} 
    \end{split}.
\end{equation*}
The elements in $\tilde{\bf{V}}(\varepsilon,\tilde{\bm{x}}^{\text{c}})$ is calculated by Eq.~\eqref{cls_rbfqr_U_calc}, and $\tilde{\bf{V}}(\varepsilon,\tilde{\bm{x}}^{\text{c}})$ has entries:
\begin{equation*}
    \tilde{\bf{V}}(\varepsilon,\tilde{\bm{x}}^{\text{c}}) = 
    \begin{bmatrix}
    \tilde{{u}}_{1}^{\text{div}}(\tilde{\bm{x}}^{\text{c}}_1) & \tilde{{u}}_{2}^{\text{div}}(\tilde{\bm{x}}^{\text{c}}_1) & \hdots & \tilde{{u}}_{N_{\text{trj}}}^{\text{div}}(\tilde{\bm{x}}^{\text{c}}_{1}) \\
    \tilde{{u}}_{1}^{\text{div}}(\tilde{\bm{x}}^{\text{c}}_2) & \tilde{{u}}_{2}^{\text{div}}(\tilde{\bm{x}}^{\text{c}}_2) & \hdots & \tilde{{u}}_{N_{\text{trj}}}^{\text{div}}(\tilde{\bm{x}}^{\text{c}}_{2}) \\
    \vdots & \vdots & \ddots & \vdots & \\
    \tilde{{u}}_{1}^{\text{div}}(\tilde{\bm{x}}^{\text{c}}_N) & \tilde{{u}}_{2}^{\text{div}}(\tilde{\bm{x}}^{\text{c}}_N) & \hdots & \tilde{{u}}_{N_{\text{trj}}}^{\text{div}}(\tilde{\bm{x}}^{\text{c}}_{N}) \\
    \end{bmatrix},
\end{equation*}
the matrices ${\bf{E}}(\varepsilon,\bm{t}^{\text{c}})$ and ${\bf{E}}_t(\varepsilon,\bm{t}^{\text{c}})$ are the same as those in Eqs.~\eqref{traj_eval} and \eqref{step1_velocity_acceleration}.

The Eqs.~\eqref{step3_ls:a} and \eqref{step3_ls:b} are established based on explicit physical intuition.
From the Lagrangian perspective, the particle  trajectory reconstructed by the RBF-QR regression (i.e., the Right-Hand Side (RHS) of Eq.~\eqref{step3_ls:a}) should be as close as possible to the modified particle locations (i.e., the Left-Hand Side (LHS) of Eq.~\eqref{step3_ls:a}), as the modified particle locations from \textit{Step~1} are the best estimates available based on the raw LPT measurement.
From the Eulerian perspective, the particle velocities along pathlines (i.e., the RHS of Eq.~\eqref{step3_ls:b}) should be equal to the divergence-free velocity field reconstructed by the constrained regression in each frame (i.e., the LHS of Eq.~\eqref{step3_ls:b}). 
Enforcing Eqs.~\eqref{step3_ls:a} and \eqref{step3_ls:b} simultaneously achieves data assimilation from both Lagrangian and Eulerian perspectives. 

Next, we solve the expansion coefficient  ${\boldsymbol{\Lambda}}$ to update particle locations.
Combining Eqs.~\eqref{step3_ls:a} and \eqref{step3_ls:b} to share the same expansion coefficient  ${\boldsymbol{\Lambda}}$, an over-determined system is established:
\begin{equation}
\bf{H} = {\boldsymbol{\Lambda}}{\bf{K}},
\label{HLambdaK}
\end{equation}
where ${\bf{K}} = [ {\bf{E}}^\text{T}(\varepsilon,\bm{t}^{\text{c}})~{\bf{E}}_{t}^\text{T}(\varepsilon,\bm{t}^{\text{c}}) ] $ and ${\bf{H}} = [ \tilde{\bf{X}}(\varepsilon,\bm{t}^\text{c})~\tilde{\bf{V}}(\varepsilon,\tilde{\bm{x}}^{\text{c}}) ]$.
The update expansion coefficient  ${\boldsymbol{\Lambda}}$ is solved by ${\boldsymbol{\Lambda}} = {\bf{H}}{\bf{K}}^{+}$, where ${\bf{K}}^{+}=({\bf{K}}^{\text{T}}{\bf{K}})^{-1}{\bf{K}}^{\text{T}}$.
The matrix ${\boldsymbol{\Lambda}}$ has dimensions of $N \times M_{\text{trj}}$ with entries:
\begin{equation}
    {\boldsymbol{\Lambda}} =
    \begin{bmatrix}
    \lambda_{1,1} & \lambda_{1,2} & \hdots & \lambda_{1,M_{\text{trj}}} \\
    \lambda_{2,1} & \lambda_{2,2} & \hdots & \lambda_{2,M_{\text{trj}}} \\
    \vdots & \vdots & \ddots & \vdots \\
    \lambda_{N,1} & \lambda_{N,2} & \hdots & \lambda_{N,M_{\text{trj}}} \\
    \end{bmatrix}.
    \label{eq: Lambda}
\end{equation}
In each row of ${\boldsymbol{\Lambda}}$, the expansion coefficients are used to approximate a trajectory for a certain particle, while in each column of ${\boldsymbol{\Lambda}}$, the expansion coefficients are used to approximate a velocity field for all particles in a certain frame.
Each row of ${\boldsymbol{\Lambda}}$ is used to update trajectories modelled by Eq.~\eqref{traj_eval} in \textit{Step~1}.
For example, for a particle $\tilde{\bm{x}}^{\text{c}}_i$, its updated trajectory is calculated by $\tilde{\bm{x}}_{i}^{\text{up}}(\varepsilon,\bm{t}^{\text{c}})={\bf{E}}(\varepsilon,\bm{t}^{\text{c}})\bm{\lambda}_i^{\text{trj}}$, where $\bm{\lambda}_i^{\text{trj}}=(\lambda_{i,1}, \lambda_{i,2}, \hdots, \lambda_{i,M_{\text{trj}}})^{\text{T}}$ is extracted from Eq.~\eqref{eq: Lambda}.
The update of particle trajectories in the $y$ and $z$ directions follow the same procedure.

The expansion coefficient matrix ${\boldsymbol{\Lambda}}$ connects the physical knowledge in both spatial and temporal dimensions. 
This is justified by the intuition that the 
Eulerian (measuring over each flow field at a certain time instant) and Lagrangian (tracking each particle over time) observations of the same flow should provide the same information. 
The shared expansion coefficient  ${\boldsymbol{\Lambda}}$ in Eq.~\eqref{HLambdaK} implies that no `discrimination' is projected to temporal and spatial dimensions, as well as to the Lagrangian and Eulerian descriptions of the flow.

\subsection{Step 4: calculate final velocity and differential quantity} 
\label{section: step 4}
\textit{Step~4} calculates the final divergence-free velocity field in each frame using the same algorithms as those in \textit{Step~2}. 
However, the updated particle locations from \textit{Step~3} and intermediate divergence-free velocities from \textit{Step~2} are used as the inputs for this step.
A spatial oversampling ratio in {\it Step~4} is defined as $\beta_2=N/M_2$.
Similar to $\beta_1$, $\beta_2$ is chosen to be larger than one.

\textit{Step~4} also computes velocity gradients.
For example, in the $x$ direction the velocity gradient at $\kappa$-th frame is given by:
\begin{equation}
    \left.\frac{\partial {\tilde{u}}(\varepsilon,\tilde{\bm{x}}^{\text{up}})} {\partial {x}} \right|_{\kappa}={\bf{E}}_{x}(\varepsilon,\tilde{\bm{x}}^{\text{up}})\bm{\lambda}_{\kappa},
\label{velocity_gradient}
\end{equation}
where $\bm{\lambda}_{\kappa}$ is the vector of the expansion coefficient in the $x$ direction.
$\bm{\lambda}_{\kappa}$ is extracted from $\bar{\bm{\lambda}}$ that is solved by Eq.~\eqref{eqn: cls_rbfqr_ex}.
${\bf{E}}_{x}(\varepsilon,\tilde{\bm{x}}^{\text{up}})$ is the RBF-QR derivative matrix based on ${\bf{E}}_{\mathcal{D}}$:
\begin{equation*}
    {\bf{E}}_{\mathcal{D}}(\varepsilon,\tilde{\bm{x}}^{\text{up}}) = E_{\mathcal{D},ij} =\psi_{\mathcal{D}} (\varepsilon,\Vert \tilde{\bm{x}}^{\text{up}}_i- \bm{x}^{\text{ref}}_j\Vert ),
\label{step2_3}
\end{equation*}
where $\bm{x}^{\text{ref}}=[\bm{x}^{\text{ref}}_1,\bm{x}^{\text{ref}}_2,\dots,\bm{x}^{\text{ref}}_{M_2}]^\text{T}$ is the spatial reference points used in \textit{Step~4}, $\bm{x}^{\text{ref}}_j=({x},{y},{z})^{\text{ref}}_j$, $j=1,2,\dots,M_2$, and $M_2$ is the number of spatial reference points in {\it Step~4}.
Velocity gradients in other directions can be calculated similarly.

When the data sets are large (e.g., more than ten thousand particles), applying the PUM is preferred.
The above velocity reconstruction is first performed in each PUM patch and then the velocity field in the entire domain is assembled using Eq.~\eqref{pum}.
The PUM settings are the same for \textit{Step~2} and \textit{Step~4} since the computational domain remains unchanged in this work.
In \textit{Step~4}, the same PUM assembly practice is applied for differential quantity fields.
As recommended by \citet{larsson2017least}, the overlap ratio of PUM patches is set to $\gamma=0.2$ to strike a balance between accuracy and computational cost.
The number of reference points in a PUM patch is chosen between about 200 and 1,000.\footnote{Less than 200 reference points per patch in 3D may be inadequate to resolve fine structures of complex flows according to our tests \citep{li2022robust} and more than 1,000 particles may lead to ill-conditioning \citep{fornberg2011stable,larsson2013stable}.}

\subsection{Spatial and temporal super-resolution} 
\label{section: superresolution}
The CLS-RBF PUM method can readily achieve spatial and temporal super-resolution.
To apply super-resolution in our method, a pseudo-particle can be placed at an arbitrary location in the domain (e.g., $(x_{\text{s}},y_{\text{s}},z_{\text{s}}) \in \Omega$) at any time instant $t$ between the first and last frames. 
In \textit{Step~1}, all modified particle locations and initial velocities are calculated at time $t$. 
Next, an intermediate divergence-free velocity field is reconstructed in \textit{Step~2}, and then the given particle locations are updated in \textit{Step~3}.
Last, the final velocities and differential quantities are calculated based on the updated particle locations from \textit{Step~3} and the intermediate velocity fields from \textit{Step~2}. 
Because the final velocity and differential quantity fields are recovered by continuous functions (i.e., the stable RBF), the velocity and velocity gradient at the location of the pseudo-particle $(x_{\text{s}},y_{\text{s}},z_{\text{s}})$ at time $t$ can be evaluated.
With this procedure, any number of pseudo-particles can be placed densely in both space and time dimensions to achieve spatiotemporal super-resolution.


\section{Result and Discussion}
\label{sec: results & discussion}

We first use synthetic LPT data generated by adding artificial noise to data from ground truth flows to test our method.
The ground truth data are time series of particle spatial coordinates, which are based on a 3D Taylor-Green vortex (TGV) \citep{taylor1937mechanism} or a Direct Numerical Simulation (DNS) of a wake behind a cylinder \citep{khojasteh2022lagrangian}.
The artificial noise was zero-mean Gaussian noise with standard deviation $\sigma$ that was proportional to the spatial span of the domain in a direction. 
For example, in the $x$ direction, the standard deviation $\sigma$ of the noise was $\sigma = \zeta L$, where $\zeta$ was the noise level and $L$ was the spatial span of the domain in the $x$ direction.
$\zeta=0.1\%$ and $\zeta=1.0\%$ was chosen in the current work to represent a medium and high noise level in an LPT experiment, respectively. 
Details of the synthetic data generation can be found in Appendix~\ref{appendix: synthetic data generation}.

We benchmark our method with some baseline algorithms. 
In the trajectory and velocity reconstruction, six algorithms, i.e., first- and second-order finite difference methods (1st and 2nd FDM), second-, third-, fourth-order polynomial regressions (2nd, 3rd, and 4th POLY), and cubic basis splines (B-splines) are used as `baselines' to compare against our method.
The baseline algorithms are briefed in Appendix~\ref{appendix: baseline algorithms}.
To assess reconstruction quality, relative errors and normalized velocity divergence are introduced. 
Relative errors ($\mathcal{E}$) of reconstructed particle spatial coordinates, velocities, and velocity gradients are quantified as:
\begin{equation*}
    \mathcal{E}=\frac{{ | \tilde{f}-f_{0} |}}{{\Vert f_{0} \Vert}_{L^{\infty}(\Omega)}} \times 100\%,
    \label{eq: error}
\end{equation*}
where $\tilde{f}$ is the reconstruction result (e.g., particle locations, velocities, and velocity gradients), and $f_0$ is the ground truth.
A normalized velocity divergence is defined as:
\begin{equation*}
    \| \nabla \cdot {\tilde{\bf{U}}} \|^{*}=\frac{\| \nabla \cdot {\tilde{\bf{U}}} \|_{L^{2}(\Omega)}} {{\left \| \left | \frac{\partial \hat{u}}{\partial x} \right | + \left| \frac{\partial \hat{v}}{\partial y} \right | + \left | \frac{\partial \hat{w}}{\partial z} \right | \right \| }_{L^{\infty}(\Omega)}},
    \label{eq: normDiv}
\end{equation*}
following \citet{luthi2002some}, where ${\tilde{\bf{U}}}$ is the reconstructed velocity vector; $\frac{\partial \hat{u}}{\partial x}$, $\frac{\partial \hat{v}}{\partial y}$ or $\frac{\partial \hat{w}}{\partial z}$ is either the ground truth (if available) or reconstructed velocity gradients.

\subsection{Validation based on Taylor-Green vortex}
\begin{figure}[!ht]
    \centering
    \includegraphics[scale=0.15]{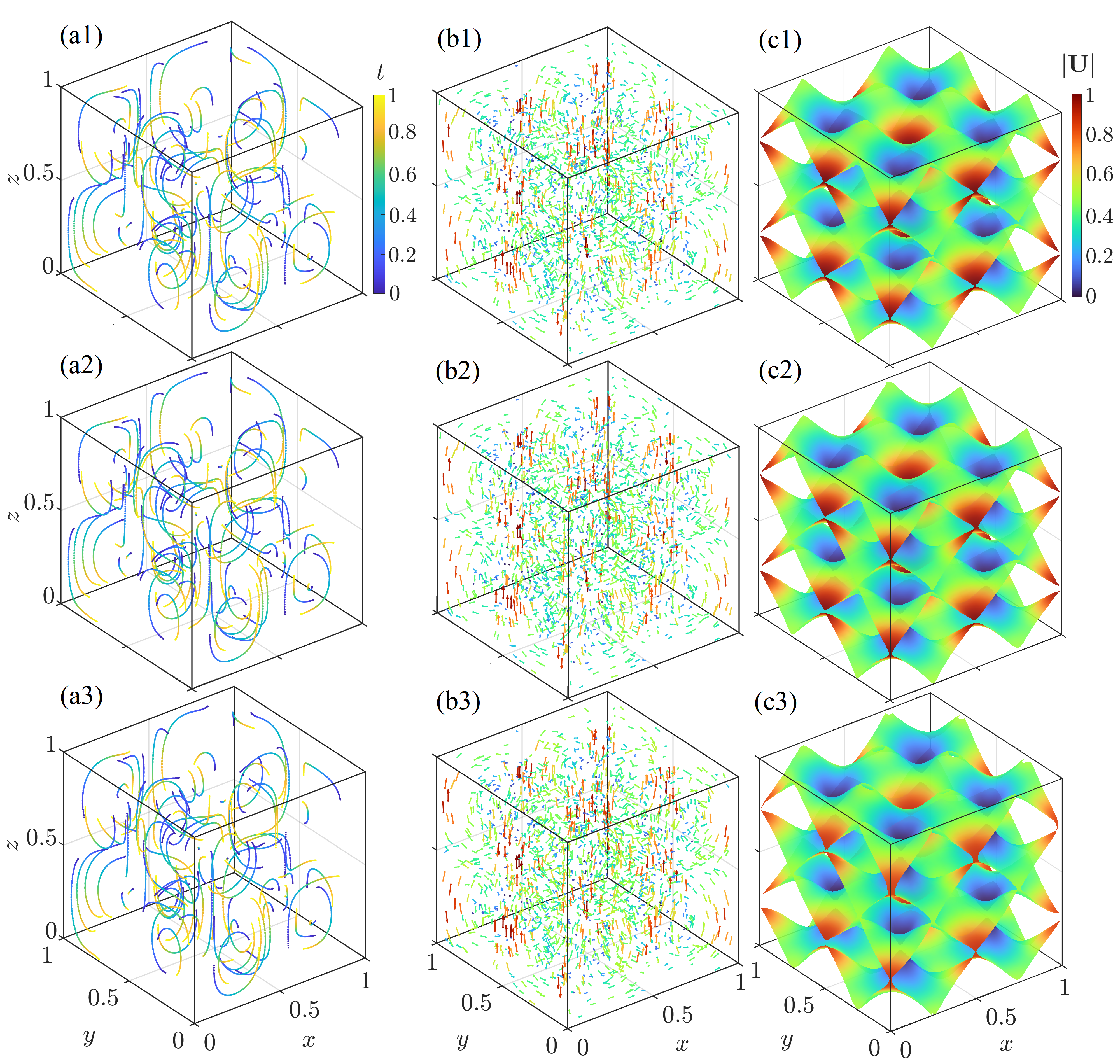}
    \caption{3D validation based on the synthetic TGV data. Left column: particle trajectories with temporal super-resolution. Middle column: particle velocity vector fields in the sixth of 11 frames. Right column: the iso-surfaces of coherent structures based on the Q-criterion (iso-value = 0.001) in the sixth of 11 frames with spatial super-resolution. Top row:  ground truth; middle and bottom rows:  reconstruction based on the synthetic data with the noise level of $\zeta \approx 0.1\%$~and~$1.0\%$, respectively. The particle trajectories are colored by time and the velocity fields and iso-surfaces are colored by the amplitude of velocity. One hundred particle trajectories and  2,000 velocity vectors out of 20,000 are shown in (a1)~--~(a3) and (b1)~--~(b3), respectively. }
    \label{fig:3DTGV_1}
\end{figure}

\begin{figure}[!th]
    \centering
    \includegraphics[scale=0.13]{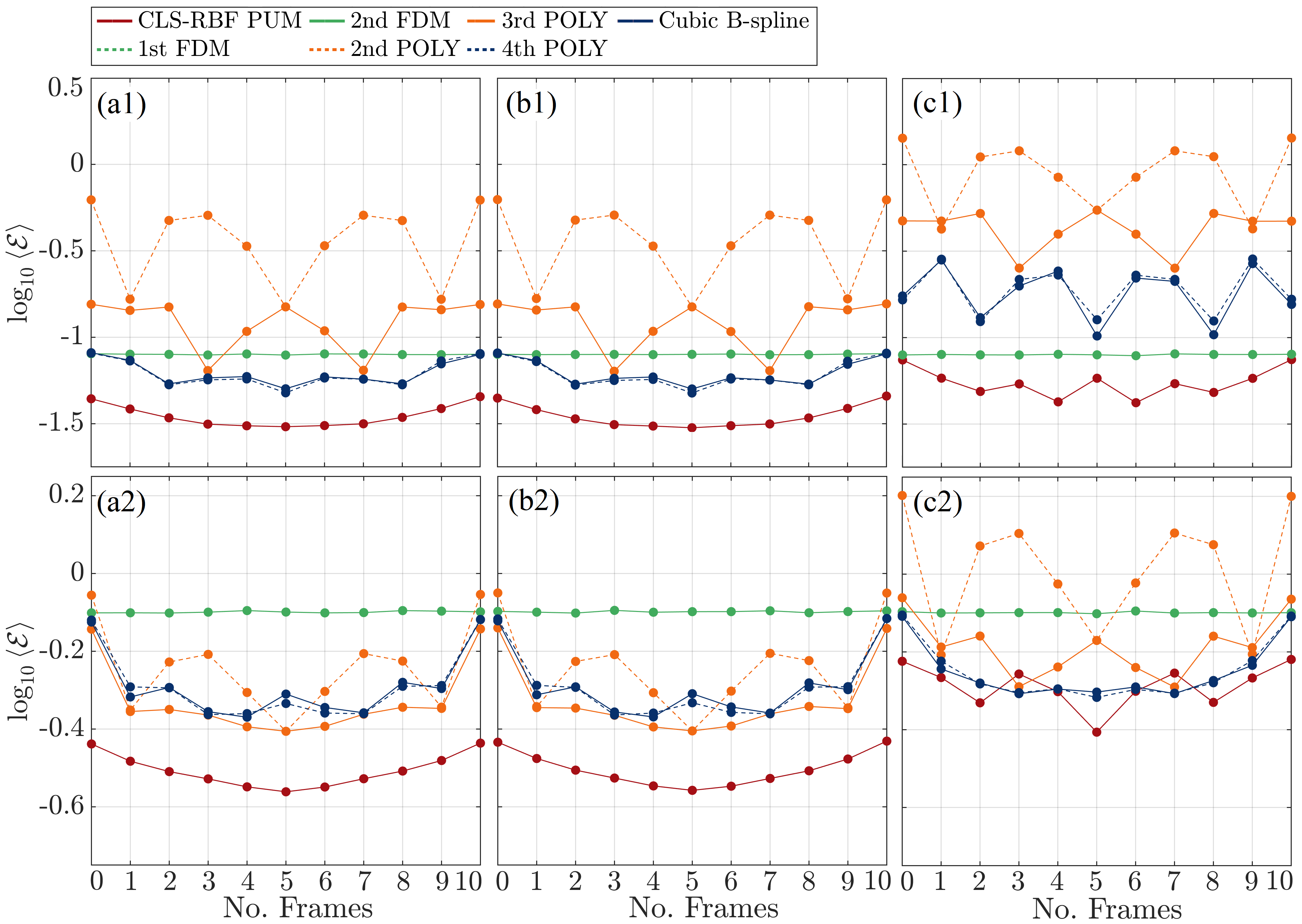}
    \caption{Reconstruction errors of particle coordinates in the TGV validation. From upper to lower rows: reconstruction based on synthetic data with 0.1\% \& 1.0\% noise level, respectively. From left to right columns: reconstruction errors in spatial coordinates of $x$, $y$, and $z$, respectively. Note that the green dashed \& solid lines are overlapped in all sub-figures because the finite difference methods only evaluate velocities. Therefore, the particle locations from the synthetic data are directly used as the trajectory outputs of the 1st \& 2nd FDMs.}
    \label{fig:3DTGV_2_coor}
\end{figure}

\begin{figure}[!ht]
    \centering
    \includegraphics[scale=0.13]{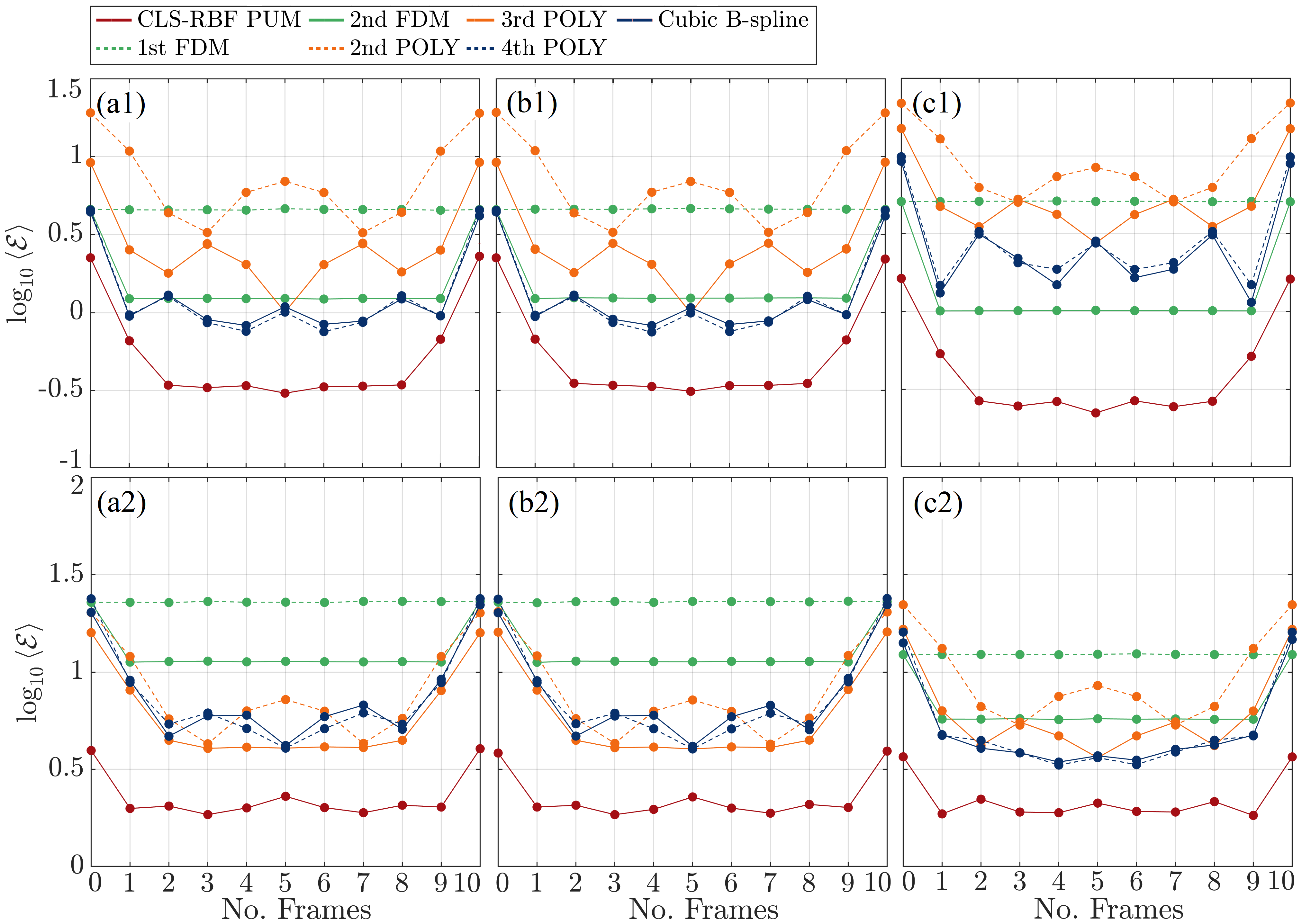}
    \caption{Reconstruction errors of particle velocities in the TGV validation. From upper to lower rows: reconstruction based on synthetic data with the noise level of $\zeta = 0.1\%$~and~$1.0\%$, respectively. From left to right columns: reconstruction errors in velocity components of $u$, $v$, and $w$, respectively.}
    \label{fig:3DTGV_2_velo}
\end{figure}

\begin{figure}[!htb]
    \centering
    \includegraphics[scale=0.15]{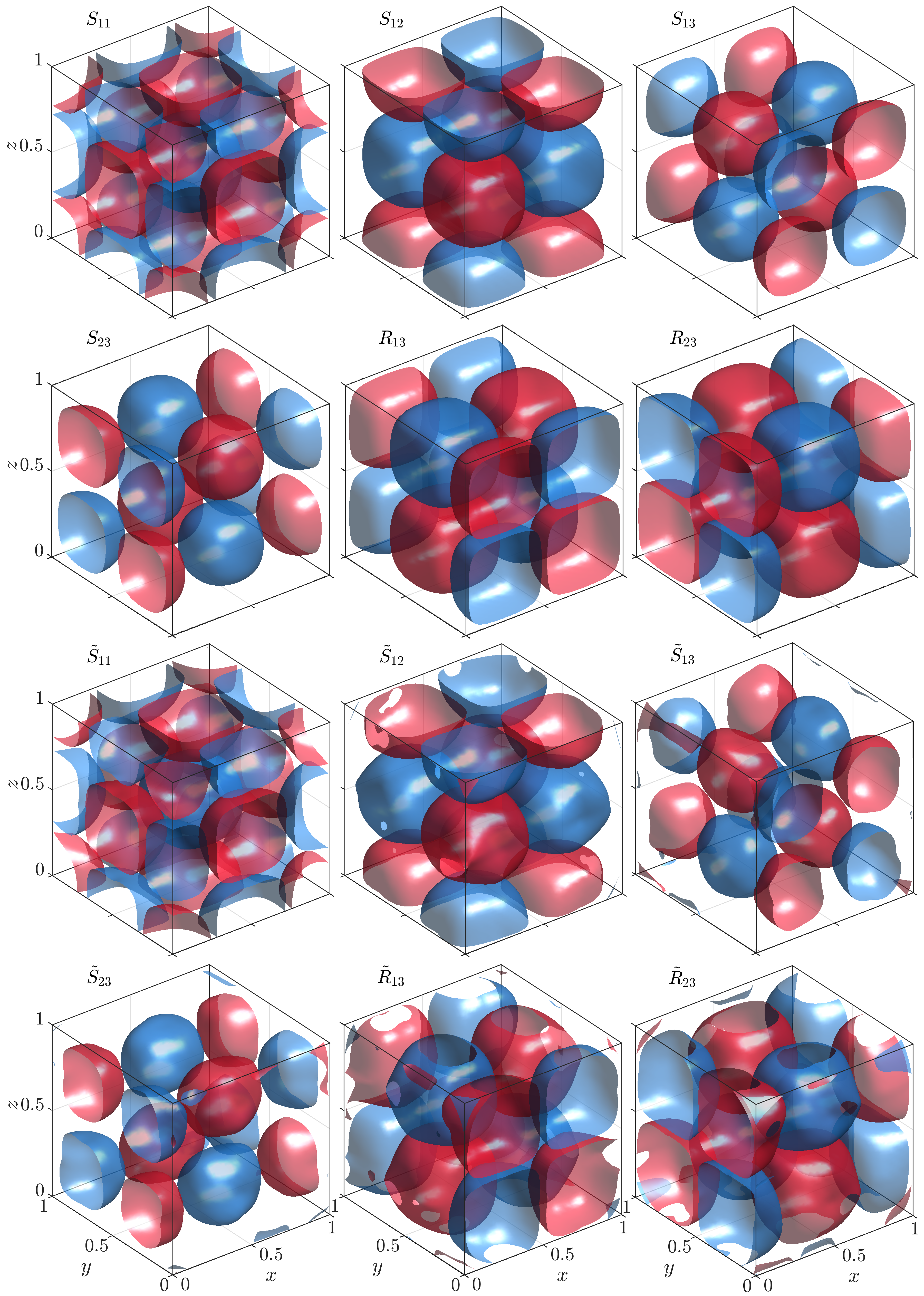}
    \caption{The iso-surfaces of strain- and rotation-rate tensors (iso-value~=~$\pm0.50$). Red and blue colors correspond to positive and negative iso-values, respectively. Upper two rows: the ground truth; lower two rows: reconstruction using the CLS-RBF PUM method. Reconstruction was in the sixth of 11 frames based on the synthetic data with 1.0\% noise added to the raw LPT data of TGV.}
    \label{fig:3DTGV_N10_3}
\end{figure}

The validation based on the synthetic TGV data is presented in Figs.~\ref{fig:3DTGV_1}~--~\ref{fig:3DTGV_N10_3}.
The TGV synthetic data have 20,000 particles scattered in the domain $\Omega$ of $x \times y \times z \in [0,1]\times[0,1]\times[0,1]$, with a time interval $\Delta t = 0.1$ between two consecutive frames.
As shown in Fig.~\ref{fig:3DTGV_1}, the reconstructed particle trajectories and velocity fields were almost identical to their ground truth, regardless of noise levels.
In Fig.~\ref{fig:3DTGV_1}c2~--~c3, only some minor distortions on the iso-surfaces appeared near the domain boundaries when high noise ($\zeta =  1.0\%$) was added to generate the synthetic data.

Figs.~\ref{fig:3DTGV_2_coor} and ~\ref{fig:3DTGV_2_velo} present the relative errors of reconstructed trajectories and velocities, respectively.
Although the reconstruction errors are relatively higher at the two ends of particle pathlines than the other frames for all methods, our method outperformed the baseline algorithms 
as the errors from our method were almost always lower than those of the baseline algorithms.
The red lines in Fig.~\ref{fig:3DTGV_2_coor}, which represent the errors in the trajectory reconstruction based on our method, lie below the green lines that denote the errors of the input data.
This evidences that our method can effectively mitigate noise in particle spatial coordinates.

Iso-surfaces of the reconstructed strain- and rotation-rate tensors based on synthetic data with high noise ($\zeta = 1
\%$) are shown in Fig.~\ref{fig:3DTGV_N10_3}.
The major structures of the flow were smooth and recognizable despite that the iso-surfaces are slightly distorted near the domain boundaries and edges (lower two rows in Fig.~\ref{fig:3DTGV_N10_3}).
The reconstructed iso-surfaces of coherent structures based on medium noise data ($\zeta =0.1\%$) were almost the same as the ground truth (not shown here for brevity).

We assess the mean and standard deviation of the relative errors in the reconstructed velocity gradients.
For noise level $\zeta=0.1\%$, the mean and standard deviation of the errors were below $2.62\%$ and $5.50\%$ for all frames, respectively. 
The errors at the two ends of the pathlines (the boundaries in time) were higher than those in the frames in between.
If the reconstruction results on the temporal boundaries were excluded, the overall reconstruction quality can be significantly improved.
For example, after excluding the first and last frames, the mean and standard deviation of the errors were below $1.38\%$ and $1.90\%$, respectively. 
For high-level artificial noise, $\zeta=1.0\%$, the mean and standard deviation of the errors were below $4.16\%$ and $4.65\%$ for all frames, respectively. 
After excluding the first and last frames, they were below $3.74\%$ and $3.91\%$, respectively.
We emphasize that the absolute value of the normalized velocity divergence is almost always below $5.7 \times 10^{-7}$ regardless of the noise level in the synthetic data.
This implies that our method effectively achieves divergence-free reconstruction.

\FloatBarrier

\subsection{Validation based on DNS of a turbulent wake}

\begin{figure}[!tbh]
    \centering
    \includegraphics[scale=0.15]{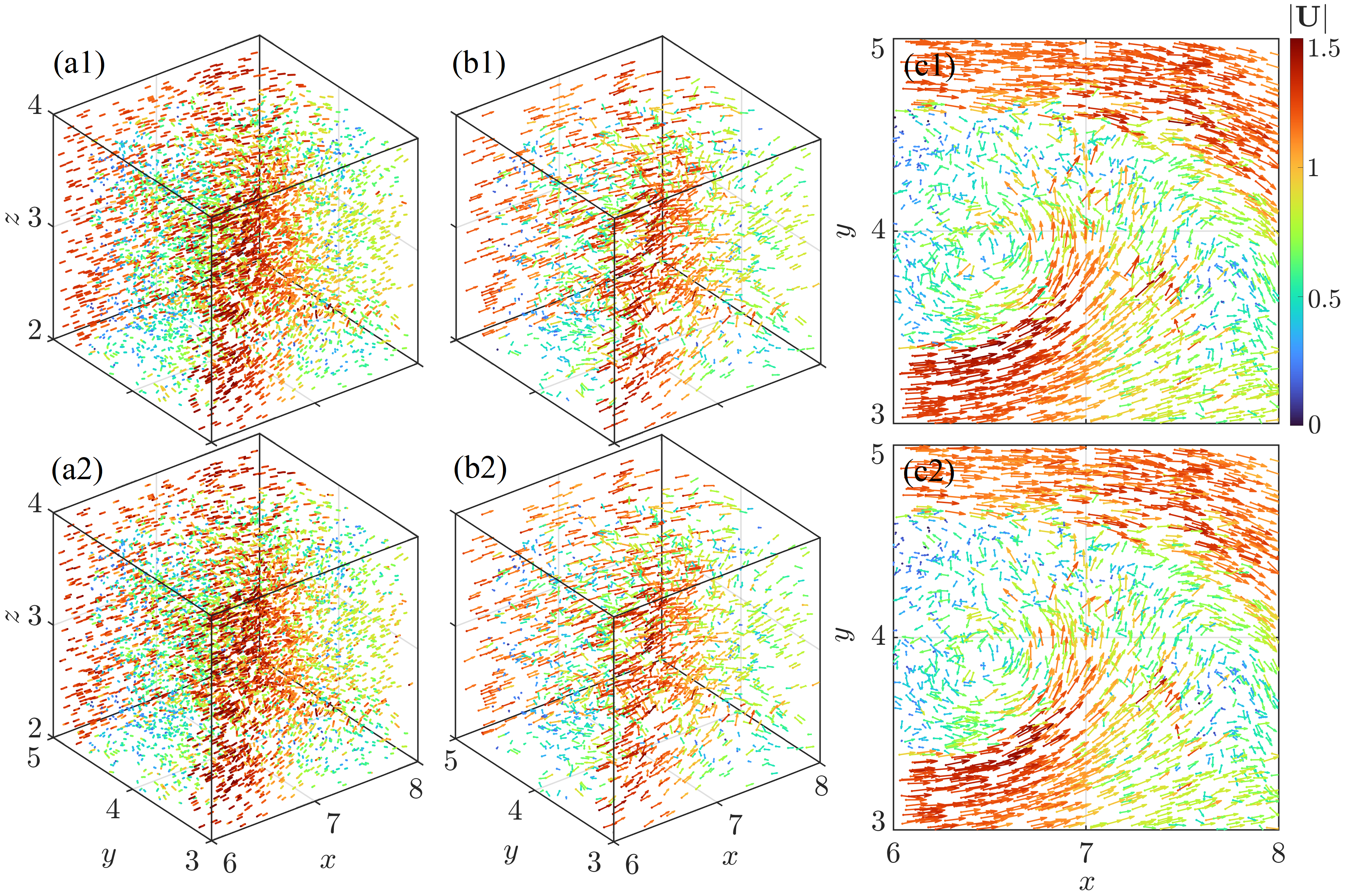}
    \caption{3D validation based on the synthetic DNS data. Upper row: the ground truth. Lower row: reconstruction based on the synthetic data with 0.1\% artificial noise added. 
    (a1)~--~(a2): particle trajectories. (b1)~--~(b2) \& (c1)~--~(c2): particle velocity fields in the sixth of 11 frames. The particle trajectories \& velocity vector fields are colored by particle velocities. 5,000 pathlines \& 2,000 velocity vectors out of 105,000 are shown in particle trajectories \& velocity fields, respectively. (c1) \& (c2) are top views from the $+z$ axis of (b1) \& (b2), respectively.}
    \label{fig:3DDNS_1}
\end{figure}

\begin{figure}[!tb]
    \centering
    \includegraphics[scale=0.14]{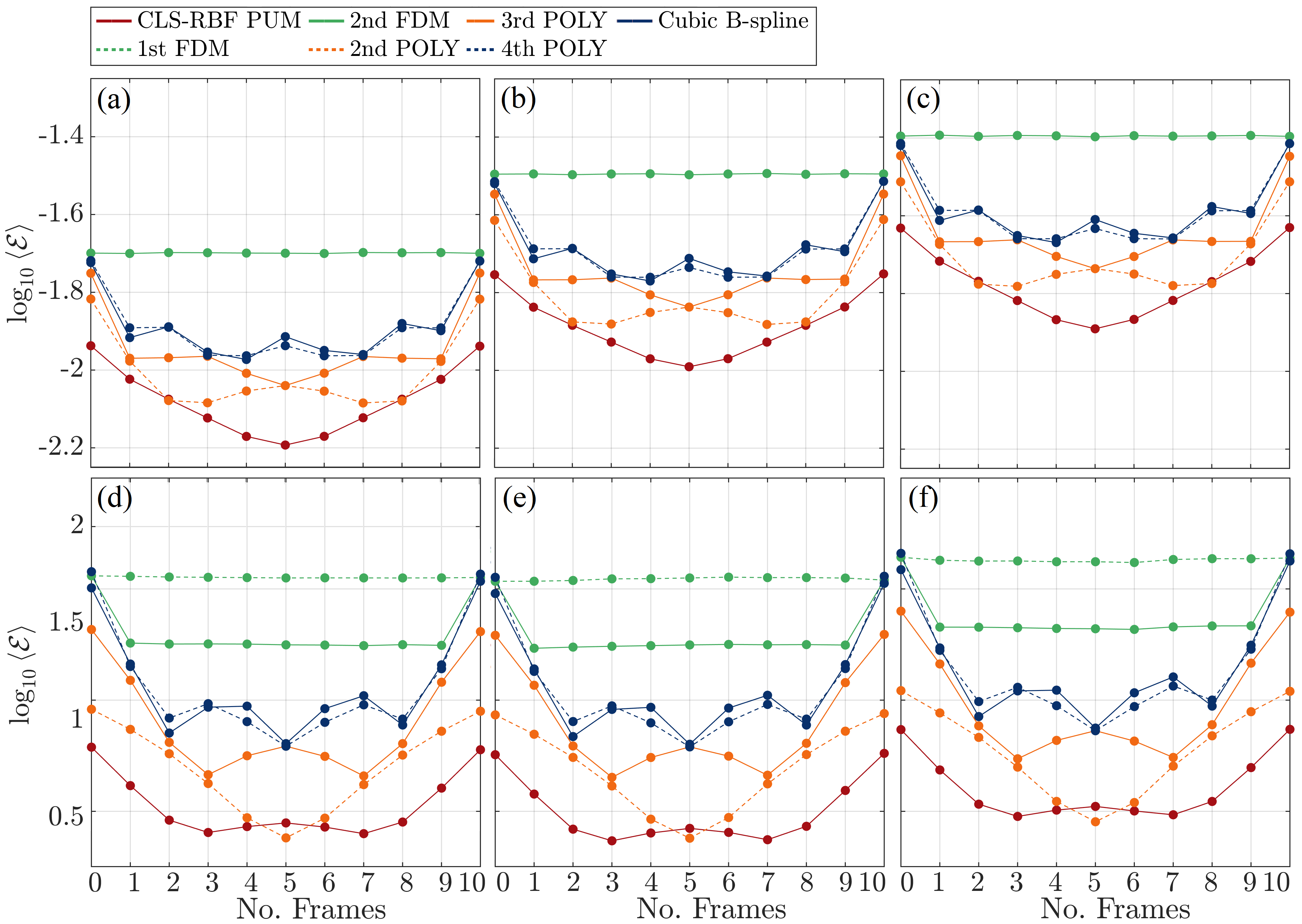}
    \caption{Reconstruction errors in the DNS validation. (a)~--~(c): reconstruction errors of spatial coordinates of $x$, $y$, and $z$ respectively; (d)~--~(f): reconstruction errors in velocity $u$, $v$, and $w$, respectively. The reconstruction is based on synthetic data with 0.1\% noise level.}
    \label{fig:3DDNS_2}
\end{figure}

The validation based on the DNS synthetic data of the turbulent wake behind a cylinder is presented in Figs.~\ref{fig:3DDNS_1} and \ref{fig:3DDNS_2}.
The DNS synthetic data have 105,000 particles scattered in the domain $\Omega$ of $x \times y \times z \in [6D,8D]\times[3D,5D]\times[2D,4D]$, with a  time interval $\Delta t = 0.00375U/D$ between two successive frames, where $U$ is free stream velocity and $D$ is the diameter of the cylinder.

As shown in Fig.~\ref{fig:3DDNS_1}, the reconstructed particle trajectories and velocity fields were almost identical to the ground truth.
As illustrated in Fig.~\ref{fig:3DDNS_2}, similar to the validation based on the TGV, the relative errors of reconstructed trajectories and velocities were almost always lower than those of the baseline algorithms. 
Note, our method significantly outperformed the baseline algorithms at the two ends of particle pathlines and suppressed noise in particle spatial coordinates.
The performance of our method on different flows (e.g., TGV or turbulent wake) with various noises ($\zeta = 0.1\%$ and $\zeta = 1.0\%$) is consistent.
The absolute value of normalized velocity divergence mostly was below $8.6 \times 10^{-5}$.

\subsection{Experimental validation based on a pulsing jet}
\label{subssection: experimental validation}
Next, we validated our method using experimental LPT data from a low-speed pulsing jet.
The experiment was conducted by \citet{sakib2022pressure} at Utah State University, US. 
The jet flow had a Reynolds number $Re_{\delta_\nu} = 400$ based on the thickness of the Stokes boundary layer. 
The experimental facility consisted of a hexagonal water tank, a cylindrical piston, an impingement plate, and optical equipment. 
The synthetic pulsing jet was generated by driving the piston using an electromagnetic shaker, and pushing the water through a circular orifice until impinged on the plate. 
A dual cavity high-speed laser illuminated the measurement area with the dimensions of $60 ~\text{mm} \times 57 ~\text{mm} \times 20~ \text{mm}$.
Four high-speed cameras recorded the jet flow and provided time-resolved LPT images.

The raw experimental data were acquired using a volumetric system and commercial software LaVision DaVis~10 (G{\"o}ttingen, Germany).
The LPT module of DaVis~10 is based on the shake-the-box~\citep{schanz2016shake} algorithm.
The velocity data were post-processed by Vortex-In-Cell\# (VIC\#) algorithm \citep{jeon2022fine} to obtain velocity gradients on a structured mesh.
About 7,100 particles with 11~frames were chosen from the original data set that had about 15,000 particles with 100 frames.\footnote{In the current experimental data, the maximum number of particles that existed in all 11 frames was about 7,100.}
This down-sampling led to a sparse data set to challenge our method. 
The reconstruction was carried out within the domain of $x \times y \times z \in [-20,30]\times[-30,25]\times[-10,10]~{\text{mm}}^3$ and the time interval between two consecutive frames was $5\times10^{-4}~{\text{s}}$.
The data reported an average of about 0.016\% nominal uncertainty in the particle spatial coordinates. 
To further assess the robustness, we tested our method on artificial error-contaminated LPT data.
These contaminated data had a noise level of $\zeta =0.2\%$. 

\begin{figure}[!hbt]
\centering
\includegraphics[scale=0.14]{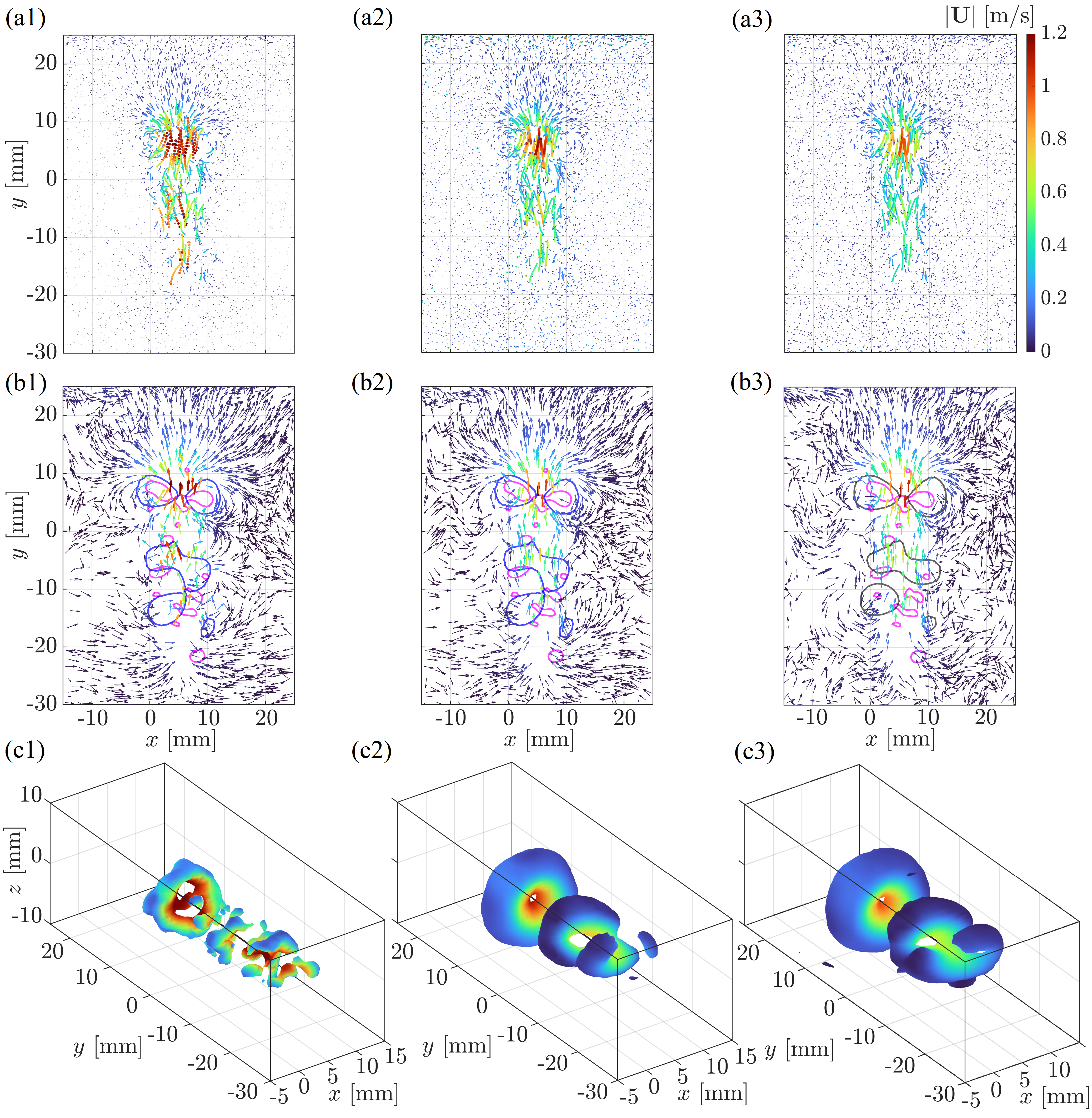}
	\caption{The experimental validation. The velocity fields and iso-surfaces are from the sixth of 11 frames. The magenta, blue, and gray contours in (b1)~--~(b3) show the reconstruction using VIC\# (iso-value = 40,000 $\text{s}^{-2}$), and our method based on raw (iso-value = 1,000 $\text{s}^{-2}$) and contaminated (iso-value = 500 $\text{s}^{-2}$) LPT data, respectively. The iso-surfaces in (c1)~--~(c3) are based on spatial super-resolution reconstruction, and their iso-values are the same as those used in (b1)~--~(b3). (a1)~--~(a3) and (b1)~--~(b3) are views from the $+z$ axis; (c1)~--~(c3) are the zoomed-in views near the jet core. Particle trajectories, velocity fields, and iso-surfaces are colored by particle velocities.}
\label{fig:3DJET}
\end{figure}

\begin{landscape}
\begin{figure}[htb]
	\centering
	\includegraphics[scale=0.15]{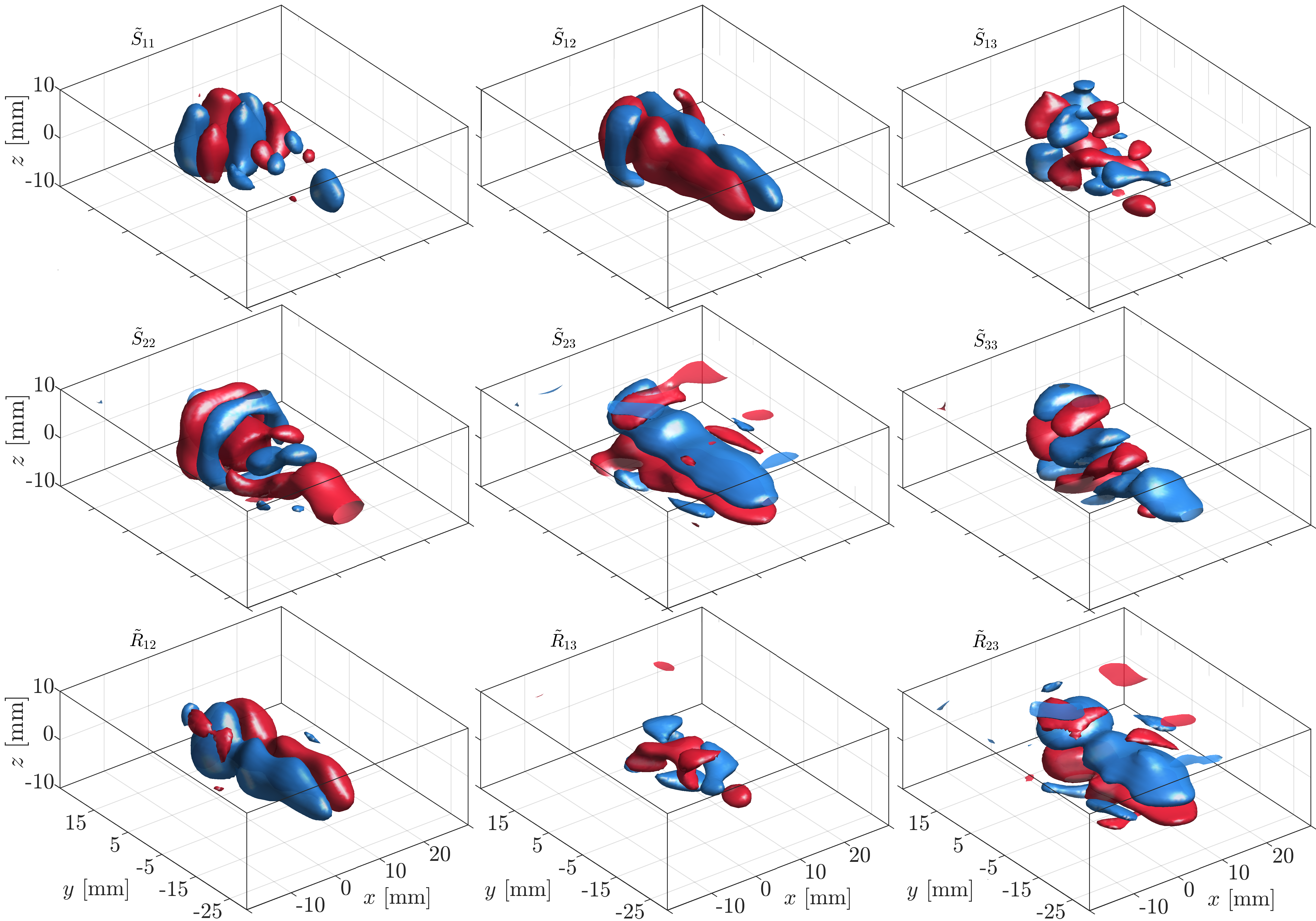}
	\caption{The iso-surfaces of reconstructed strain- \& rotation-rate tensors in the sixth of 11 frames using the CLS-RBF PUM method (iso-value = $\pm15$ $\text{s}^{-1}$). Red and blue colors correspond to positive and negative iso-values, respectively. Only nine components are shown here.}
\label{fig:3DJET_Tensor}
\end{figure}
\end{landscape}

The experimental validation results are shown in Figs.~\ref{fig:3DJET} and \ref{fig:3DJET_Tensor}.
In Fig.~\ref{fig:3DJET}, the left column (i.e., Fig.~\ref{fig:3DJET}a1~--~c1) presents reconstructions obtained from DaVis 10. 
The central and right columns illustrate the CLS-RBF PUM reconstructions based on raw and contaminated LPT data, respectively.
The top row (i.e., Fig.~\ref{fig:3DJET}a1~--~a3) shows reconstructions of particle trajectories. The virtual size of particles is proportional to their velocities for visualization purposes.
The middle row illustrates reconstructed velocity fields. 
We only visualized the velocity vectors within the range of $z \in [-3,+3]$~mm, which covered the jet core. 
To emphasize the directions of the particle velocities, the length of the vectors was normalized and projected on the $z=0$ plane. 
The bottom row represents iso-surfaces of coherent structures based on the Q-criterion.
Intersections between the iso-surfaces and $z=0$ plane are contours identifying the vertical region of the flow sliced at $z=0$, and are overlaid on the quiver plots in the middle row (i.e., Fig.~\ref{fig:3DJET}b1~--~b3).

As shown in Fig.~\ref{fig:3DJET}a2~--~a3, smooth particle trajectories were  recovered by our method, whose profiles resembled those obtained from DaVis~10 (see Fig.~\ref{fig:3DJET}a1), regardless of particle coordinates being significantly contaminated by the artificial noise (see Fig.~\ref{fig:3DJET}a3). 
In addition, two trailing jets (at $y \approx -5$ and $-15$~mm) were revealed in the wake of the leading pulsing jet (at $y \approx 5$~mm). 
Note, our method was able to reconstruct trajectories with temporal super-resolution, where each pathline consists of 51 frames (see Fig.~\ref{fig:3DJET}a2~--~a3).

Compared Fig.~\ref{fig:3DJET}b2~--~b3 with Fig.~\ref{fig:3DJET}b1, the velocity fields reconstructed by our method were virtually smoother, in terms of the transition of the vector directions over the space. 
This observation is more apparent near the core of the trailing jets.
Furthermore, our method effectively captured major structures of the flow (illustrated by blue and gray contours in Fig.~\ref{fig:3DJET}b1~--~b3, and the coherent structures in Fig.~\ref{fig:3DJET}c2~--~c3).
We can observe three vortical structures associated with the leading pulse jet and the two trailing jets (see e.g., Fig.~\ref{fig:3DJET}a1).
The normalized velocity divergence was below $3.13 \times 10^{-7}$ in all frames for our method (e.g., Fig.~\ref{fig:3DJET}b2~--~b3), regardless of the added artificial noise or not. 
This exhibits that the divergence-free constraint was enforced.
On the contrary, the VIC\# had a normalized velocity divergence of about $3.18$.

Fig.~\ref{fig:3DJET}c2~--~c3 represents the reconstructed coherent structures based on the Q-criterion. 
One large and two small toroidal structures emerged in the domain and their locations corresponded to the three high-speed areas observed in Fig.~\ref{fig:3DJET}a1~--~a3.
This agreement between the coherent structures and the particle trajectory and velocity data further validates our method. 
Comparing the middle and right columns in Fig.~\ref{fig:3DJET}, despite the overwhelming artificial noise added to the down-sampled data, no discernible difference was observed between the two reconstruction results.
This indicates that our method is robust on noisy sparse data.

The iso-surfaces of strain- and rotation-rate tensors are illustrated in Fig.~\ref{fig:3DJET_Tensor}. 
They show the kinematics of fluid parcels that may not be visually apparent from the velocity or vorticity fields alone. 
As examples, we interpret two components of the strain-rate tensor.  
For $\tilde{S}_{12}$, two major tubular structures with reversed colors adhering to each other, developed along the $y$ axis, and two minor vortical structures warped the major ones.
The reversed colors of the major structures indicate shear deformations near the jet core, which were caused by the radial velocity gradients of the jet. 
The wavy tubular structure with three bumps reflects the leading pulsing jet and the two trailing ones.
The two minor structures suggest that fluid parcels experienced shear deformations in opposite directions compared to the closest major structure, generating reversed flows that brought fluid parcels back to the jet.
Regarding $\tilde{S}_{13}$, the staggering pattern parallel to the $x$-$z$ plane indicates the shear deformation of fluid parcels in the front and back of the core of the leading jet.
The fluid parcels at the leading front plane of the vortex rings tended to be elongated along the circumferential direction and shrunk along the radial directions, while the fluid parcels just behind them underwent reversed deformations. 
This explains the forward movement of the vortex rings.
Noting that the front patterns were larger than the back ones, one can tell that the vortex ring was expanding by observing this single frame.

\section{Conclusion}
\label{sec: conclusion}
In this paper, we propose the CLS-RBF PUM method, a novel 3D divergence-free Lagrangian flow field reconstruction technique. 
It can reconstruct particle trajectories, velocities, and differential quantities (e.g., pressure gradients, strain- and rotation-rate tensors, and coherent structures based on the Q-criterion) from raw Lagrangian Particle Tracking (LPT) data.
This method integrates the Constrained Least Squares (CLS), a stable Radial Basis Function (RBF-QR), and Partition-of-Unity Method (PUM) into one comprehensive reconstruction strategy.
The CLS serves as a platform for LPT data regression and enforcing physical constraints.
The RBF-QR approximates particle trajectories along pathlines, using the time as an independent variable, and approximates velocity fields for each frame, with the particle spatial coordinates being their independent variables. 
The PUM localizes the reconstruction to enhance computational efficiency and stability. 

The intuition behind the CLS-RBF PUM method is straightforward. 
By assimilating the velocity field reconstructed based on Lagrangian and Eulerian perspectives, we intrinsically incorporate the information in the temporal and spatial dimensions with physical constraints enforced to improve flow field reconstruction and offer several advantages.
This method directly reconstructs flow fields at scattered data points without Lagrangian-Eulerian data conversions and can achieve super-resolution at any time and location and enforce physical constraints. 
The constraints are velocity solenoidal conditions for incompressible flows in the current work while accommodating other constraints as needed.
Large-scale LPT data sets with a substantial number of particles and frames can be efficiently processed and parallel computing is achievable.
It demonstrates high accuracy and robustness, even when handling highly contaminated data with low spatial and/or temporal resolution.

The tests based on synthetic and experimental LPT data show the competence of the CLS-RBF PUM method.
Validation based on synthetic data has exhibited the superior trajectory and velocity reconstruction performance of our method compared to various baseline algorithms.
The tests based on a pulsing jet experiment further confirm the effectiveness of our method.
In summary, the CLS-RBF PUM method offers a versatile solution for reconstructing Lagrangian flow fields based on the raw LPT data with accuracy, robustness, and physical constraints being satisfied, and can be the foundation of other downstream post-processing and data assimilation.

\section*{Acknowledgments}
We thank Prof. Elizabeth Larsson at Uppsala University for discussing the RBF-QR and sharing codes with us.
We are also thankful to Dr. Md Nazmus Sakib at Utah State University for conducting experiments and providing data and timely support. 
This project is partially supported by the Natural Sciences and Engineering Research Council of Canada (NSERC).

\clearpage
\section*{Appendices}
\begin{appendices}
\section{Synthetic Data}
\label{appendix: synthetic data generation}
\subsection{Taylor-Green vortex (TGV)}
\label{subsubsection: TGV synthetic}
To generate the TGV synthetic data, we adopted a velocity field:
\begin{equation}
\begin{split}
    u = & \alpha_1 \cos{(\omega (x-d_x))}\sin{(\omega (y-d_y))}\sin{(\omega (z-d_z))} \\
    v = & \alpha_2 \sin{(\omega (x-d_x))}\cos{(\omega (y-d_y))}\sin{(\omega (z-d_z))} \\
    w = & \alpha_3 \sin{(\omega (x-d_x))}\sin{(\omega (y-d_y))}\cos{(\omega (z-d_z))} 
\end{split},
\label{tgv_3d}
\end{equation}
where $\alpha_1=\alpha_2=0.5$, $\alpha_3=-1$, and $\omega=2 \pi$, 
$d_x=d_y=d_z=0.25$ were the offsets of the TGV flow structure in the $x$, $y$ and $z$ directions, respectively. 
In the first snapshot, 20,000 particles were randomly placed in the domain $\Omega$ of $x \times y \times z \in [0,1]\times[0,1]\times[0,1]$.
The particle locations in the next snapshot were calculated by the forward Euler method: $\bm{x}_{\kappa+1}=\bm{x}_{\kappa}+\delta t {\bf{U}}_{\kappa}$, where $\bm{x}_{\kappa}=({x}_{\kappa},{y}_{\kappa},{z}_{\kappa})$ and $\bm{x}_{\kappa+1}$ were the particle locations in the $\kappa$-th and $(\kappa+1)$-th snapshots, respectively;
${\bf{U}}_{\kappa}=(u_{\kappa},v_{\kappa},w_{\kappa})$ was the velocity calculated by Eq.~\eqref{tgv_3d} in the $\kappa$-th snapshot;
$\delta t$ was the time interval between snapshots and here we chose $\delta t=5 \times 10^{-5}$.
After the integration, particle trajectories with $2 \times 10^{4}$ snapshots were down-sampled to 11 frames to generate the ground truth of the raw LPT data.
Artificial noise (see Sect.~\ref{sec: results & discussion} for details) was added to the particle spatial coordinates of the TGV-based ground truth.

\subsection{Direct Numerical Simulation (DNS)}
\label{subsubsection: DNS synthetic}
For the validation based on the DNS, synthetic data were generated by adding artificial noise to an available Lagrangian data set.
More details about the data set can be found in \citet{khojasteh2022lagrangian} and we briefly summarize here. 
The DNS was conducted using the \verb+Incompact3D+ solver \citep{laizet2009high} to simulate a flow with a free stream velocity $U$ passing a cylinder with a diameter $D$ at a Reynolds number of $Re=3,900$.

In the Lagrangian data sets, about 200,000 particles were scattered in the domain.
The particle velocities were calculated by tri-linear interpolations using the nearest eight nodal data in space.
The particle trajectories were integrated by a fourth-order Runge-Kutta method.
The synthetic data had 105,000 particles that were down-sampled from the Lagrangian data set.
These particles were distributed in a wake region, located $0.5D$ downstream of the cylinder, with the dimensions of $x \times y \times z \in [6D,8D]\times[3D,5D]\times[2D,4D]$. 
A total of 11 frames were selected from 350 DNS frames and these 11 frames had a time interval $\Delta t=0.00375U/D$. 
The noise was defined the same way as that used in the TGV validation, with $\zeta=0.1\%$ to represent a typical noise in LPT data.

\section{Baseline Algorithms}
\label{appendix: baseline algorithms}
\subsection{Finite difference methods}
We use the velocity component $u$ calculation as an example. 
In a given frame, the 1st and 2nd order finite difference methods (FDMs) calculate velocities by the forward Euler method (i.e., Eq.~\eqref{eq: fdm_1}) and central difference method (i.e., Eq.~\eqref{eq: fdm_2}), respectively:
\begin{subequations}
\begin{align}
        \tilde{u}(t_\kappa) &= \frac{\hat{x}_{\kappa+1}-\hat{x}_{\kappa}}{\Delta{t}}, 
        \label{eq: fdm_1} \\
        \tilde{u}(t_\kappa) &= \frac{\hat{x}_{\kappa+1}-\hat{x}_{\kappa-1}}{2\Delta{t}},
        \label{eq: fdm_2}
\end{align}
\end{subequations}
where $\hat{x}_{\kappa}$ is the particle coordinate along the $x$ direction in the $\kappa$-th frame from LPT experiments, $\Delta{t}$ is the time interval between two consecutive frames, and $t_\kappa$ is the time instant in the $\kappa$-th frame.
The velocities of the first and last frames are calculated by the first-order finite difference method.
Since the FDMs only evaluate particle velocities, the particle locations that are `output' from the FDMs are assumed to be the same as those in the synthetic data,
e.g., $\tilde{x}(t_\kappa)=\hat{x}(t_\kappa)$.

\subsection{Polynomial regression}
We use trajectory and velocity reconstruction in the $x$ direction as an example.
First, the trajectory polynomial model function $\tilde{x}(t)$ is given by $\tilde{x}(t)=\sum^m_{j=0}{p_{m,j} t^j}$ where $p_{m,j}$ is the polynomial coefficient, $m$ is the order of polynomials, and $t$ is the time.
For instance, the trajectory and velocity model functions of the 2nd polynomial (2nd-POLY) are
\begin{equation*}
\begin{split}
    \tilde{x}(t)=p_{2,2} t^2+ p_{2,1}t+p_{2,0} \\
    \tilde{u}(t)=\dot{\tilde{x}}(t)=2p_{2,2} t+ p_{2,1}
\end{split}.
\end{equation*}
The trajectory and velocity approximation functions of the other polynomials are similar to those in the $x$ direction of the 2nd POLY.

Next, we calculate the polynomial coefficient $p_{m,j}$.
A residual $\mathcal{R}$ between measured particle locations $\hat{x}_{\kappa}$ and the polynomial model function $\tilde{x}({t}_{\kappa})$ is minimized  $    \min~ \mathcal{R}=\sum^{N_{\text{trj}}}_{\kappa=1}\Vert \tilde{x}({t}_{\kappa})-\hat{x}_{\kappa} \Vert^2,$
where ${t}_{\kappa}$ is the time instant in the $\kappa$-th frame.
Setting the gradient of the residual $\mathcal{R}$ regarding $p_{m,j}$ to zero (i.e., $\partial{\mathcal{R}}/\partial{p_{m,j}} =0$) and the polynomial coefficient $p_{m,j}$ can be explicitly solved.
The polynomial coefficients in the $y$ and $z$ directions are the same as that in the $x$ direction.
Once all the coefficients are retrieved, the polynomial model functions can be constructed and used to approximate particle trajectories, velocities, and accelerations.

\subsection{Cubic B-splines}
The cubic B-spline algorithm is based on the MATLAB built-in functions from the Curve Fitting Toolbox\texttrademark.
We adopt a function \verb+spap2(piece,k,x,y)+ with \verb+piece=2+ and \verb+k=4+ to create a cubic spline with two pieces joined together, bypass setting knots.
\verb+x+ and \verb+y+ are the given data points and their values, respectively.
Then we use a function \verb+fnval(f,xe)+, in which \verb+f+=\verb+spap2(piece,k,xe,y)+ is the spline function calculated above, and \verb+xe+ is the evaluation points.
To calculate first-order derivatives for velocity as an example, we use \verb+fnder(f,d)+, where \verb+d=1+.

\section{Differential Quantity}
\label{appendix: differential quantity computation}
\subsection{Strain-rate and rotation-rate tensor}
\label{subsubsection: strain- and rotation-rate tensors}
Strain-rate and rotation-rate tensors are kinematics quantities that describe the rate of change of a fluid parcel regarding deformation and rotation, respectively.
The strain-rate tensor ${\bm{S}}$ and rotation-rate tensor ${\bm{R}}$ are the symmetric and anti-symmetric parts of the velocity gradient $\nabla {\bf{U}}$, respectively: $\nabla {\bf{U}} =\frac{1}{2}(\mathit{U}_{i,j}+\mathit{U}_{j,i})+\frac{1}{2}(\mathit{U}_{i,j}-\mathit{U}_{j,i})$, 
${\bm{S}} = S_{ij} =\frac{1}{2}(\mathit{U}_{i,j}+\mathit{U}_{j,i})$, and $ {\bm{R}} = R_{ij} =\frac{1}{2}(\mathit{U}_{i,j}-\mathit{U}_{j,i})$.
Once all components of  $\mathit{U}_{i,j}$ are calculated in \textit{Step~4} of the CLS-RBF PUM method, ${\bm{S}}$ and ${\bm{R}}$ can be evaluated.

\subsection{Coherent structure based on Q-criterion}
\label{subsubsection: coherent and q}
 The Q-criterion is a vortex identification method first proposed by \citet{hunt1988eddies} to identify vortical structures in the flow.
The Q-criterion is given by \citep{hunt1988eddies,haller2005objective}:
\begin{equation}
    Q=\frac{1}{2}(\Vert {\bm{R}}^2 \Vert - \Vert \bm{S}^2 \Vert)>0
    \label{Q_criterion}
\end{equation}
where $\bm{R}$ is the rotation-rate tensor, and $\bm{S}$ is the strain-rate tensor. 
After ${\bm{S}}$ and ${\bm{R}}$ are calculated, vortices can be found using Eq.~\eqref{Q_criterion}. 
Other criteria can be achieved similarly.

\end{appendices}

\bibliographystyle{apalike}
\bibliography{reference}
\end{document}